\documentclass[fleqn,usenatbib]{mnras}

\usepackage{newtxtext,newtxmath}
\usepackage[T1]{fontenc}
\usepackage{ae,aecompl}

\usepackage{graphicx}	% Including figure files
\usepackage[export]{adjustbox} % Allow right justification of figures
\usepackage{amsmath}	% Advanced maths commands
\usepackage{amssymb}	% Extra maths symbols

\newcommand{\bvec}[1]{\mbox{\boldmath ${#1}$}}

\title[Sun-as-a-star radial-velocity observations]{Three years of Sun-as-a-star radial-velocity observations on the approach to solar minimum.}

\author[A. Collier Cameron et al.]{
A.\,Collier Cameron,$^{1,15,19}$\thanks{E-mail: acc4@st-andrews.ac.uk}
A.\,Mortier,$^{2,1,15}$
D.\,Phillips,$^{3}$
X.\,Dumusque,$^{4}$
\newauthor
R.\,D.\,Haywood,$^{3,17}$
N.\,Langellier,$^{3,5}$
C.\,A.\,Watson,$^{6}$ %- coauthorship accepted 23/10/2018
H.\,M.\,Cegla,$^{4,18}$
J.\,Costes,$^{6}$ %- coauthorship accepted 23/10/2018
\newauthor
D.\,Charbonneau,$^{3}$
A. Coffinet,$^{4}$
D.\,W.\,Latham,$^{3}$
M.\,Lopez-Morales,$^{3}$
L.\,Malavolta,$^{7,8}$
\newauthor
J.\,Maldonado,$^{9}$
G.\,Micela,$^{9}$
T.\,Milbourne,$^{3,5}$
E.\,Molinari,$^{10}$ %- coauthorship accepted 23/10/2018
S.\,H.\,Saar,$^{3}$
S.\,Thompson,$^{2}$
\newauthor
N. Buchschacher,$^{4}$
M.\,Cecconi,$^{11}$
R.\,Cosentino,$^{11}$
A.\,Ghedina,$^{11}$
A.\,Glenday,$^{3}$
\newauthor
M. Gonzalez,$^{11}$
C.-H.\,Li,$^{3}$
M.\,Lodi,$^{11}$
C.\,Lovis,$^{4}$
F.\,Pepe,$^{4}$
E.\,Poretti,$^{11,12}$
K.\,Rice,$^{13,16}$ %- coauthorship accepted 25/10/2018
\newauthor
D.\,Sasselov,$^{3}$ %- coauthorship accepted 24/10/2018
A.\,Sozzetti,$^{14}$
A.\,Szentgyorgyi,$^{3}$
S.\,Udry$^{4}$ 
and
R.\,Walsworth$^{3,5}$
\\
\\
$^{1}$SUPA, School of Physics and Astronomy, University of St Andrews, North Haugh, St Andrews KY16 9SS, UK\\
$^{2}$Astrophysics Group, Cavendish Laboratory, J.J. Thomson Avenue, Cambridge CB3 0HE, UK\\
$^{3}$Harvard-Smithsonian Center for Astrophysics, 60 Garden Street, Cambridge, MA 02138, USA\\
$^{4}$Observatoire Astronomique de l'Universit\'{e} de G\'en\`eve, 51 Chemin des Maillettes, 1290 Sauverny, Suisse\\
$^{5}$Department of Physics, Harvard University, 17 Oxford Street, Cambridge MA 02138, USA\\
$^{6}$Astrophysics Research Centre, School of Mathematics and Physics, Queen's University Belfast, University Road, Belfast, BT7 1NN, UK\\
$^{7}$INAF - Osservatorio Astronomico di Padova, Vicolo dell'Osservatorio 5, 35122 Padova, Italy\\
$^{8}$Dipartimento di Fisica e Astronomia ``Galileo Galilei", Universita' di Padova, Vicolo dell'Osservatorio 3, I-35122 Padova, Italy\\
$^{9}$INAF - Osservatorio Astronomico di Palermo, Piazza del Parlamento 1, 90134 Palermo, Italy\\
$^{10}$INAF - Osservatorio Astronomico di Cagliari, via della Scienza 5, 09047, Selargius, Italy\\
$^{11}$INAF - Fundaci\'on Galileo Galilei, Rambla Jos\'e Ana Fernandez P\'erez 7, E-38712 Bre\~na Baja, Tenerife, Spain\\
$^{12}$INAF - Osservatorio Astronomico di Brera, Via E. Bianchi 46, 23807 Merate (LC), Italy\\
$^{13}$SUPA, Institute for Astronomy, Royal Observatory, University of Edinburgh, Blackford Hill, Edinburgh EH93HJ, UK\\
$^{14}$INAF - Osservatorio Astrofisico di Torino, via Osservatorio 20, 10025 Pino Torinese, Italy\\
$^{15}$Centre for Exoplanet Science,  University of St Andrews,  St Andrews,  UK\\
$^{16}$Centre for Exoplanet Science,  University of Edinburgh,  Edinburgh,  UK\\
$^{17}$NASA Sagan Fellow\\
$^{18}$CHEOPS Fellow, SNSF NCCR-PlanetS\\
$^{19}$Visiting Scientist, Lowell Observatory, 1400 Mars Hill Rd, Flagstaff, AZ 86001, USA \\
}

\date{Accepted 2019 April 26. Received 2019 April 25; in original form 2018 November 01}

\pubyear{2019}

\begin{document}
\label{firstpage}
\pagerange{\pageref{firstpage}--\pageref{lastpage}}
\maketitle

\begin{abstract}
The time-variable velocity fields of solar-type stars limit the precision of radial-velocity determinations of their planets' masses, obstructing detection of Earth twins. Since 2015 July we have been monitoring disc-integrated sunlight in daytime using a purpose-built solar telescope and fibre feed to the HARPS-N stellar radial-velocity spectrometer. We present and analyse the solar radial-velocity measurements and cross-correlation function (CCF) parameters obtained in the first 3 years of observation, interpreting them in the context of spatially-resolved solar observations. We describe a Bayesian mixture-model approach to automated data-quality monitoring. We provide dynamical and daily differential-extinction corrections to place the radial velocities in the heliocentric reference frame, and the CCF shape parameters in the sidereal frame. We achieve a photon-noise limited radial-velocity precision better than 0.43 m s$^{-1}$ per 5-minute observation. The day-to-day precision is limited by zero-point calibration uncertainty with an RMS scatter of about 0.4 m~s$^{-1}$. We find significant signals from granulation and solar activity. Within a day, granulation noise dominates, with an amplitude of about 0.4 m~s$^{-1}$ and an autocorrelation half-life of 15 minutes. On longer timescales, activity dominates. Sunspot groups broaden the CCF as they cross the solar disc. Facular regions temporarily reduce the intrinsic asymmetry of the CCF. The radial-velocity increase that accompanies an active-region passage has a typical amplitude of 5 m~s$^{-1}$ and is correlated with the line asymmetry, but leads it by  3 days. Spectral line-shape variability thus shows promise as a proxy for recovering the true radial velocity.
\end{abstract}


%%%%%%%%%%%%%%%%%%%%%%%%%%%%%%%%%%%%%%%%%%%%%%%%%%

%%%%%%%%%%%%%%%%% BODY OF PAPER %%%%%%%%%%%%%%%%%%

\section{Introduction}

%{\em Motivation for HARPS-N solar telescope project. Refer to Dumusque, Phillips papers.}

The Sun is the only star that can be observed with spatial resolution fine enough to discern the finest convective and magnetic elements that decorate its surface. Tiny though they are, these small surface elements have a profound impact on the integrated solar spectrum. The contrast between the hot, upwelling cores of granules only a few hundred km across and the cooler surrounding downflow lanes gives rise to global spectral-line asymmetries \citep{1981A&A....96..345D}. These asymmetries are strongly suppressed by small-scale magnetic fields in the faculae that surround large sunspot groups \citep{2013ApJ...763...95C}. The finite number of granules, and their finite lifetimes, give rise to statistical fluctuations in the global solar radial velocity \citep{2006A&A...445..661L}. The changing filling factor of sunspots and faculae as the Sun rotates gives rise to larger perturbations in global radial velocity and line asymmetries \citep{2010A&A...512A..39M}.

Early campaigns to monitor the solar radial velocity in integrated sunlight (e.g. \citealt {1987ApJ...316..771D}) were motivated by the need to understand the intrinsic variability of solar-type stars at a time when high-precision studies of stellar radial velocities (e.g. \citealt{1979PASP...91..540C}) were in their infancy. At about the same time, the Birmingham Solar Oscillations Network (BiSON, \citealt{1996SoPh..168....1C}) started a 39-year campaign of radial-velocity monitoring of integrated sunlight, with the goal of using low-order $p$-modes in the solar oscillation spectrum to probe the Sun's deep interior.

In the era of ultra-high precision radial velocity instruments such as ESO's High-Accuracy Radial-velocity Planet Searcher (HARPS, \citealt{2004A&A...423..385P}) and its northern counterpart HARPS-N \citep{2012SPIE.8446E..1VC}, Keck's High Resolution \'{E}chelle Spectrometer (HIRES, \citealt{1994SPIE.2198..362V}), the Echelle SPectrograph for Rocky Exoplanets and Stable Spectroscopic Observations (ESPRESSO, \citealt{2014AN....335....8P}) at the VLT, 
the  EXtreme PREcision Spectrometer (EXPRES, \citealt{2016SPIE.9908E..6TJ}) at Lowell Observatory and others following up transiting terrestrial-sized planet candidates from the {\em CoRoT}, {\em Kepler/K2} and {\em TESS} space photometry missions, the need has become acute to understand the frequency spectrum and origins of stellar radial-velocity variability. Studies such as the recent ``radial-velocity fitting challenge" of \cite{2017A&A...598A.133D} are designed to test the efficacy of novel analysis tools for modeling star-induced RV variability as part of the measurement process, and have highlighted the need to understand the underlying physics.

Using an approach developed by \cite{2000A&A...353..380F} for modelling solar irradiance variations, \cite{2010A&A...512A..39M} pioneered the study of the effects of different types of solar activity on the global solar radial velocity. By partitioning solar images from the Michaelson Doppler Imager (MDI) aboard SoHo into quiet-sun, sunspot and facular regions according to their continuum brightness and magnetic flux density, and isolating the relative velocities of the different components in the Dopplergrams, Meunier et al. concluded that the dominant contributor to solar radial-velocity (RV) variability is convective suppression of the granular blueshift. This prediction was confirmed observationally by \cite{2016MNRAS.457.3637H}, who used the same technique on images from the Helioseismic and Magnetic Imager (HMI) on the Solar Dynamics Observatory (SDO) to model radial-velocity variations in integrated sunlight reflected from asteroid 4/Vesta.

The HARPS-N solar telescope \citep{2016SPIE.9912E..6ZP,2015ApJ...814L..21D} was conceived with a longer-term goal in mind: to monitor precisely the solar radial velocity during the day using the same instrument as is being used at night for measuring the reflex orbital motions of exoplanet host stars. By using the stellar data-reduction pipeline for the solar spectra, we aim to characterise the impact of both solar variability and instrumental and data-reduction systematics on the radial velocities delivered by the instrument.

The analysis of the solar data is complicated by two considerations that do not apply to stellar targets. The observer and the target are both participants in solar-system gravitational dynamics, and the Sun is not a point source. The purpose of the present paper is to describe the methods used to correct for these non-stellar effects, transforming the data to the equivalent of the sidereal, heliocentric frame. The goal is to separate the effects of solar photospheric physics from those of solar-system dynamics and differential atmospheric effects.

In Section~\ref{sec:observations} we describe briefly the instrument, the  observing strategy and the data-reduction pipeline. We develop a Gaussian mixture-model approach to determine daily extinction coefficients and to quantify the reliability of data points affected by short-term obscuration of parts of the solar disc. We correct the radial velocities for differential extinction across the solar rotation profile and use the good data within single days to assess the level of residual $p$-mode and granulation noise on minutes-to-hours timescales. In Section~\ref{sec:corrections} we transform the radial velocities delivered by the pipeline from the barycentric frame (which is dominated by the synodic radial-velocity signal of Jupiter) to the heliocentric frame. We provide algorithms for correcting line-profile moments of the HARPS-N cross-correlation function (CCF) for the Earth's changing orbital velocity and the obliquity of the solar rotation axis to the ecliptic. In Section~\ref{sec:trends} we analyse the behaviour of the radial velocity and CCF profile parameters on timescales from days to years. We identify fluctuations in the width and asymmetry of the CCF profile with the passages of sunspot groups and large facular regions across the solar disc, respectively. We identify 
correlations and temporal offsets between these parameters and the radial velocity, and discuss their viability as proxy indicators for disentangling exoplanetary orbital reflex motion from host-star activity.

\section{HARPS-N solar telescope observations}
\label{sec:observations}

%{\em Brief summary of instrument configuration, daily coverage, cadence, calibration.
%i.e. all standard HARPS-N methodology. Start date, end date, daily coverage, cadence.}

\subsection{Instrument and observing strategy}
\label{sec:instrument} % used for referring to this section from elsewhere

The HARPS-N solar telescope comprises a small guided telescope on an amateur mount, housed in a perspex dome on the exterior of the enclosure of the 3.58-m Telescopio Nazionale Galileo (TNG) at the Observatorio del Roque de los Muchachos, Spain. Its 7.6-cm achromatic lens of 200mm focal length feeds sunlight via an integrating sphere and an optical fibre into the calibration unit of the HARPS-N spectrograph. The telescope, fibre feed and control systems are described by \cite{2016SPIE.9912E..6ZP}. Early results from the instrument were published by \cite{2015ApJ...814L..21D}, confirming that it achieves uniformity of throughput better than 1 part in $10^4$over the solar disc. This performance is essential to achieve velocity precision of 10 cm s$^{-1}$ across the width of the solar rotation profile.

\begin{figure}
\includegraphics[width=\columnwidth]{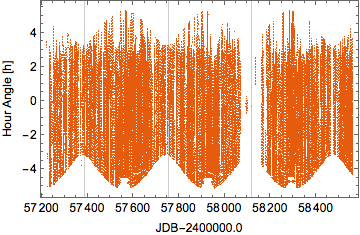}
\caption{Hour angle coverage for all observations satisfying quality-control criteria described in this paper from the start of operations in 2015 July up to the end of 2018 October. Vertical gridlines denote calendar-year boundaries at the start of 2016, 2017 and 2018. The 3-month gap in usable observations in late 2017 and early 2018 arose through damage to the fibre coupling the solar telescope to the HARPS-N calibration unit. The early-morning delay to the start of observations around midsummer occurs because the Sun rises behind part of the TNG enclosure.}
\label{fig:hrangl}
\end{figure}

The instrument observes the Sun continuously each day, with 5-minute integration times designed to average out solar $p$-mode oscillations. The seasonal variations in hour angle are shown in Fig.~\ref{fig:hrangl}. Observing begins each clear day when the Sun rises above an altitude of $\sim 20$ degrees, and ends either at the same altitude or earlier, if the instrument is required for afternoon setup by the night-time observer.

\subsection{HARPS-N data reduction pipeline}
\label{sec:pipeline} % used for referring to this section from elsewhere

The data are reduced using the same HARPS-N Data-Reduction System (DRS) employed for night-time stellar radial velocimetry. 
Calibration exposures taken after the end of solar observations and before the start of night-time observing each day provide order-by-order information on the locations of the echelle orders and the wavelength calibration scale.
Fabry-Perot exposures are recorded simultaneously with the solar exposures to monitor instrumental drift, via a second optical fibre from the spectrograph calibration unit. Following optimal extraction \citep{1986PASP...98..609H, 1989PASP..101.1032M} to obtain one-dimensional background-subtracted spectra in each order, the data are calibrated in wavelength. They are then cross-correlated with a digital mask \citep{1996A&AS..119..373B,2002A&A...388..632P} derived from a typical G2 stellar spectrum, and corrected for instrumental drift derived from the Fabry-Perot spectrum. The radial velocity is computed as the mean velocity of a Gaussian fit to the CCF profile. The (dimensionless) contrast of the CCF is expressed as the maximum depth of the fitted Gaussian, expressed as a percentage of the surrounding pseudo-continuum. The area $W$ of the fitted Gaussian is then proportional to the product of the contrast $C$ and the full width at half maximum depth ($F\equiv{\rm FWHM}$):
\begin{equation}
W=\frac{C F}{2}\sqrt{\frac{\pi}{\ln{2}}}. 
\end{equation}
The units of $W$ and $F$ are the same, in this case velocities in km~s$^{-1}$, because the pseudo-continuum of the CCF is normalised to unity. Although $C$ is normally expressed as a percentage, the expression above treats it as a fraction $0<C<1$. The ``area'' is therefore defined in a manner similar to the equivalent width of a single spectral line.

The CCF itself is not perfectly symmetric, mainly because of the convective asymmetry in solar and stellar line profiles resulting from the brightness and velocity structure of the photospheric granulation pattern. This asymmetry is quantified as the difference in the line-bisector velocity in the upper and lower parts of the CCF, commonly referred to as the Bisector Inverse Slope (BIS; \citealt{2001A&A...379..279Q}). 

The DRS computes the formal uncertainty in each radial velocity by propagating the photon-noise error of the spectrum through all stages of extraction and cross-correlation, into the CCF. The error of the RV precision is then measured using the derivative of the CCF \citep{2001A&A...374..733B}.
The median radial-velocity precision achieved with a standard 5-minute exposure is 0.43 ms $^{-1}$ in observing conditions of high transparency; the upper 95th percentile is 0.77 m s$^{-1}$. 

\begin{figure*}
\includegraphics[width=2\columnwidth]{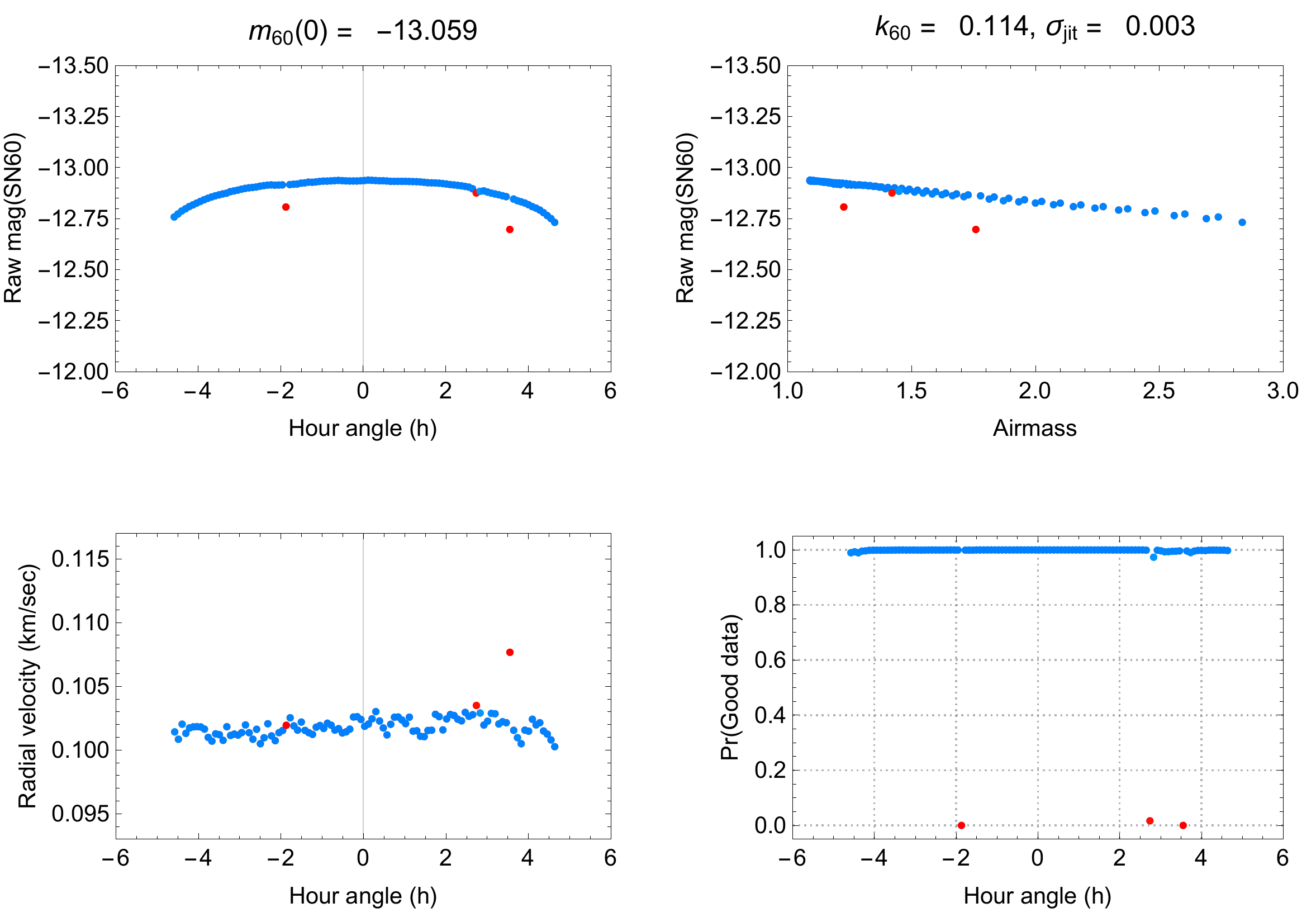}
\caption{The upper panels show the instrumental magnitude in order 60 as functions of hour angle (left) and airmass (right) on a clear day (2018 Apr 03) with good transparency but intermittent clouds or guiding errors during the day. The lower panels show the radial velocity corrected for differential extinction, and the median probability for each data point that it belongs to the foreground (good) mixture population as given by Eq.~\ref{eq:marginal} of Sec.~\ref{sec:dataquality}: Pr(good data) is synonymous with $p(q_i=0|y)$}. Points are colour-coded from blue to red in order of descending median probability that they belong to the foreground population. The extinction coefficient $k_{60}$ and transparency fluctuation amplitude $\sigma_{\rm jit}$ are small.
\label{fig:qualplot_clear}
\end{figure*}

\subsection{Data quality assessment}
\label{sec:dataquality} % used for referring to this section from elsewhere

%{\em Identification of frames affected by guiding errors and cloud. Determination of daily extinction coefficient and zero point. Flux drop-outs, changes in RV, FWHM, BIS due to cloud.}

Under normal circumstances the SNR of the radial velocity is determined primarily by photon shot noise and wavelength calibration error. Large systematic errors in the radial velocity  can, however, arise when part of the solar disc is obscured, for example by cloud or by temporary tracking errors in the telescope mount. This happens because the solar disc has a finite angular diameter and a projected equatorial rotation speed of order 2 km s$^{-1}$. Partial obscuration of the solar disc distorts the rotation profile and corrupts the measured radial velocity. It is therefore mandatory to identify and flag data points with anomalously low fluxes relative to neighbouring points. 

The worst outliers are eliminated by rejecting the 5 percent of the data with the poorest estimated radial-velocity errors, followed by an iterative 6-sigma clip of the radial velocities in the heliocentric frame. 

It is relatively straightforward to perform more rigorous automated data-quality assessment because the spectral fluxes observed in cloud-free conditions should follow an exponential extinction law as a function of airmass. To determine the apparent magnitude of the Sun, we use the SNR estimate for pipeline order 60 (echelle order 98, central wavelength 6245 \AA) recorded in the HARPS-N data headers. The SNR is proportional to the square root of the recorded photon count. We therefore construct a sequence of instrumental magnitudes of which the $i$th measurement is

\begin{equation}
y_i \equiv m_{60,i}=-5\log_{10}SN_{60,i},
\end{equation}
with corresponding magnitude uncertainty
\begin{equation}
\sigma_i = \frac{2.5}{\ln 10}\frac{1}{SN_{60,i}}.
\end{equation}

On a day of near-optimal observing conditions, such as the clear spring day affected by occasional scattered clouds illustrated in Fig.~\ref{fig:qualplot_clear}, we expect the relation between the instrumental magnitude $m_{60}(x_i)$ in order 60 and the airmass $x_i$ of the centre of the solar disc to be represented by a linear (Bouguer's Law, \citealt{1729edos.book.....B}) model
\begin{eqnarray}
m_{60}(x_i)&=&m_{60}(x=0)+k_{60} x_i\nonumber\\
\Rightarrow m_{60}(x_i)-m_{60}(\hat{x})&=&k_{60}(x_i-\hat{x}).
\end{eqnarray}
The slope $k_{60}$ of the linear extinction law is the primary extinction coefficient, while $m_{60}(x=0)$ is the instrumental magnitude extrapolated to zero airmass using the extinction law. We use the inverse variance-weighted mean airmass $\hat{x}$ and the corresponding magnitude $m_{60}(x=\hat{x})$ as the origin for the linear regression, to eliminate correlation between the slope and the zero-point of the extinction law. 

\begin{figure*}
\includegraphics[width=2\columnwidth]{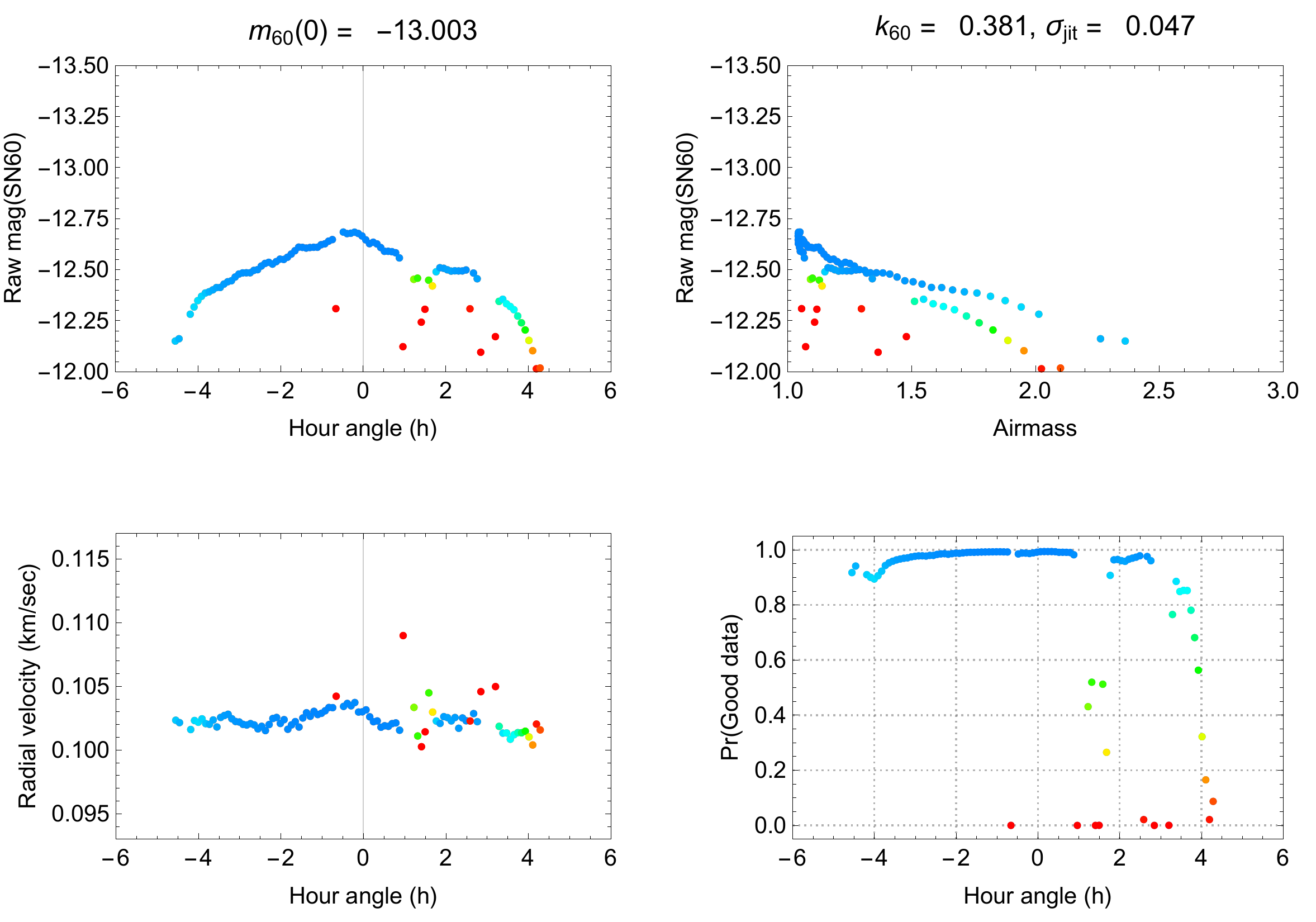}
\caption{As for Fig.~\ref{fig:qualplot_clear}, showing data obtained on 2017 Aug 20, a summer day affected by poor atmospheric transparency caused by the Saharan dust conditions known as {\em calima}. The extinction coefficient and transparency fluctuations are high, but the spatial variations in extinction are smooth enough to yield reliable velocities except when scattered clouds cause severe decreases in flux.}
\label{fig:qualplot_calima}
\end{figure*}

In addition to the fiducial magnitude $m_{60}(\hat{x})$ at the mean airmass and the extinction coefficient $k_{60}$ we include a white-noise variance parameter $\sigma^2_{\rm jit}$ representing low-level transparency fluctuations. This mitigates the risk of rejecting usable data in observing conditions where the transparency is imperfect but slowly-varying, as is the case shown in Fig.~\ref{fig:qualplot_calima} for a day affected by the {\em calima} conditions often encountered in summer when fine dust is blown westward from the Sahara desert over the observatory. Even on a clear day, the extinction coefficient may vary subtly with time and/or hour angle \citep{1993A&A...268..369P}. 

It is not clear that simple methods such as iterative sigma-clipping can make a reliable distinction between good and bad data under {\em calima} conditions in particular. Instead, we adopt a more rigorous and robust
Bayesian mixture-model approach \citep[e.g.][]{2010arXiv1008.4686H} to identify those data points that lie close enough to the daily linear extinction law to be considered reliable, and to distinguish them from a separate background population of outliers that do not. For this purpose we introduce a discrete classifier $q_i$ which takes the value 0 for a good (``foreground'') observation, and 1 for a ``background'' outlier.

The extinction model yields the likelihood that a single measurement $y_i=m_{60}(x_i)$  was obtained in good (i.e. $q_i=0$) conditions:
\begin{equation}
p(y_i | x_i, \sigma_i,\theta,q_i=0)=\frac{1}{\sqrt{2\pi(\sigma^2_i+\theta^2_3)}}
\exp\left(-\frac{[y_i-\theta_1-\theta_2(x_i-\hat{x})]^2}{2(\sigma^2_i+\theta^2_3)}\right),
\label{eq:prgood}
\end{equation}
where we have defined a vector of extinction-model parameters $\bvec{\theta}$ such that $\theta_1\equiv m_{60}(\hat{x})$, $\theta_2\equiv k_{60}$, and $\theta_3\equiv\sigma_{\rm jit}$.

Outliers produced by clouds or guiding errors generally lie below the linear relation, and are treated as having been drawn from a distinct ``background" statistical population with mean $b_{\rm bg}$ (relative to the inverse variance-weighted mean $\hat{y}$) and variance $\sigma^2_{\rm bg}$.

For a point drawn from the background (i.e. $q_i=1$) outlier population, and an extended vector of model parameters $\bvec{\theta}\equiv\{m_{60}(\hat{x}),k_{60},\sigma_{\rm jit},b_{\rm bg},\sigma_{\rm bg}\}$ such that $\theta_4=b_{\rm bg}$ and $\theta_5=\sigma_{\rm bg}$, the likelihood is
\begin{equation}
p(y_i | x_i, \sigma_i,\theta,q_i=1)=\frac{1}{\sqrt{2\pi(\sigma^2_i+\theta^2_5)}}
\exp\left(-\frac{[y_i-\hat{y}-\theta_4]^2}{2(\sigma^2_i+\theta^2_5)}\right).
\label{eq:prbad}
\end{equation}

As \cite{2010arXiv1008.4686H} and \cite{2014zndo.15856} point out, the marginal likelihood for the full dataset can be expressed as
\begin{equation}
p(\bvec{y} | \bvec{x},\bvec{\sigma},\bvec{\theta})
=\prod_{i=1}^N p(y_i | x_i, \sigma_i,\bvec{\theta})
\end{equation}
where $N$ is the number of observations on the day in question, and
\begin{equation}
p(y_i | x_i, \sigma_i,\bvec{\theta})=\sum_{q_i} p(q_i) p(y_i | x_i, \sigma_i,\bvec{\theta},q_i).
\end{equation}
The simple prior $p(q_i=0) = Q$ and $p(q_i=1) = 1-Q$, yields the likelihood function 
\begin{eqnarray}
p(\bvec{y} | \bvec{x},\bvec{\sigma},\bvec{\theta})=\prod_{i=1}^N &[&Q p(y_i | x_i, \sigma_i,\bvec{\theta},q_i=0)\nonumber\\
&&+(1-Q)p(y_i | x_i, \sigma_i,\bvec{\theta},q_i=1)]
\label{eq:likely}
\end{eqnarray}
\citep{2010arXiv1008.4686H}. 

In the present application, the prior $Q$ can be thought of as the fraction of all observations obtained on a given day that belong to the foreground (``good'') population. It is a quantity that must be derived from the data themselves, because weather conditions change from day to day.

\begin{figure*}
\includegraphics[width=2\columnwidth]{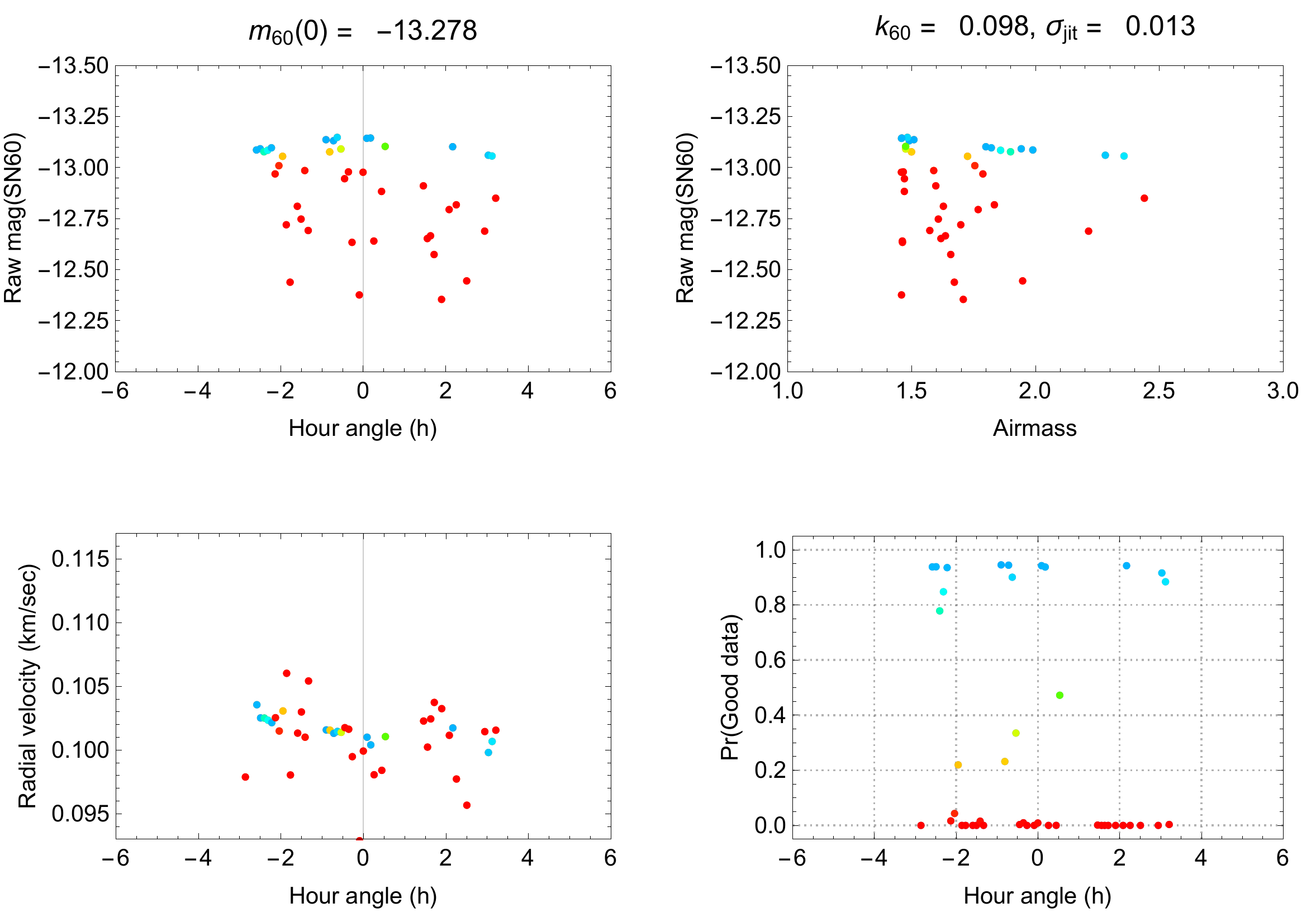}
\caption{As for Fig.~\ref{fig:qualplot_clear}, showing data obtained on 2016 Jan 29, a winter day affected by widespread variable cloud cover.  The extinction coefficient and transparency fluctuations are low during the occasional intervals when the sky is clear, but the majority of points show depressed fluxes and strong scatter more characteristic of the background population.}
\label{fig:qualplot_cirrus}
\end{figure*}

\begin{table*}
\caption[]{Bayesian prior probability distributions for the parameters of the Gaussian mixture model used to determine daily extinction coefficients and data quality.}
\label{tab:priors}
\begin{tabular}{ccc}
\hline\\
Variable & Parameter & Prior \\
\hline\\
$\theta_1$ & Fiducial magnitude $m_{60}(\hat{x})$ at airmass $x=\hat{x}$ &  $\mathcal{U}[-0.4,0.4]$\\
$\theta_2$ & Extinction coefficient $k_{60}$ & $\mathcal{U}[0.075, 0.4]$ \\
$\theta_3$ & Transparency fluctuation amplitude $\sigma_{\rm jit}$& $\mathcal{U}[0.001, 0.08]$ \\
$\theta_4$ & Outlier zero point offset $b_{\rm bg}$ relative to $\hat{y}$& $\mathcal{U}[0.0,0.8]$ \\
$\theta_5$ & Outlier standard deviation $\sigma_{\rm bg}$& $\mathcal{U}[0.1,0.7]$ \\
$Q$ & Fraction of data in foreground population & $\mathcal{U}[0.0,1.0]$ \\
\hline\\
\end{tabular}
\end{table*}

We sample from the distribution described by Eq.~\ref{eq:likely} using a simple Markov-chain Monte Carlo scheme, saving the values of $p(y_i | x_i, \sigma_i,\theta,q_i=0)$ and $p(y_i | x_i, \sigma_i,\theta,q_i=1)$ for each sample for later use. This allows us to estimate the conditional foreground probabilities marginalised over the model parameters via the method described by \cite{2014zndo.15856}:
\begin{equation}
p(q_i | \bvec{y})\approx\frac{1}{M}\sum_{m=1}^M p(q_i | \bvec{y},\bvec{\theta}^{(m)}),
\label{eq:marginal}
\end{equation}
where $M$ is the number of samples in the Markov chain. In this expression, the conditional probability that a given data point $i$ is good ($q_i=0$) or bad ($q_i=1$) for a specified parameter set $\theta$ is given by
\begin{equation}
p(q_i | \bvec{y},\bvec{\theta})=\frac{p(q_i)p(\bvec{y} | \bvec{x}, \bvec{\sigma},\bvec{\theta},q_i)}
{Q p(\bvec{y} | \bvec{x}, \bvec{\sigma},\bvec{\theta},q_i=0)+(1-Q)p(\bvec{y} | \bvec{x}, \bvec{\sigma},\bvec{\theta},q_i=1)}.
\end{equation}

As before, the priors $p(q_i)$ take values $Q$ and $1-Q$ for $q_i=0$ and 1 respectively, while
\begin{equation}
p(\bvec{y} | \bvec{x}, \bvec{\sigma},\bvec{\theta},q_i=0)
=\prod_{i=1}^N p(y_i | x_i,\sigma_i,q_i=0)
\end{equation}
using Eq.~\ref{eq:prgood} (and similarly for $q_i=1$ using Eq.~\ref{eq:prbad}).

The marginalised mixture membership probabilities described by Eq.~\ref{eq:marginal} (referred to hereafter as ``mixture probabilities'') provide a robust way of quantifying data quality. Fig.~\ref{fig:qualplot_cirrus} illustrates the performance of the model on a day of generally poor and variable transparency, probably caused by cirrus cloud. Points lying along the upper envelope of the airmass plot in the upper-right panel follow a linear extinction law, indicating that they were obtained in clear intervals between cloud passages. As in Figs.~\ref{fig:qualplot_clear} and \ref{fig:qualplot_calima}, the radial velocities of the observations classified as having mixture probabilities greater than 0.9 of being ``good'' (blue) show a much lower point-to-point scatter than those with lower mixture probabilities (cyan, green, yellow, red). This confirms that the mixture-model approach is successful, and that a threshold probability of about 0.9 is appropriate for detecting and filtering out observations with anomalous extinction high enough to bias the radial-velocity measurements.

The prior probability distributions for the parameters are listed in Table~\ref{tab:priors}. To ensure orthogonality between the fiducial magnitude $\theta_1=m_{60}(\hat{x})$ and the extinction coefficient $\theta_2=k_{60}$, we subtract the inverse variance-weighted mean airmass $\hat{x}$ from all airmass values, and the corresponding fiducial magnitude $m_{60}(\hat{x})$ from all magnitude values. The extinction coefficient was allowed to vary between 0.075 mag/airmass, which represents a transparency slightly better than  is encountered in any of the 904 daily datasets examined, and 0.4 mag/airmass, representing the worst transparency encountered in {\em calima} conditions that can still yield a meaningful airmass plot. Short-term transparency fluctuations with a standard deviation as great as 0.08 mag/airmass are allowed in the foreground population; anything worse than this is likely to give differential extinction across the solar disc bad enough to distort the radial velocity measurements. The mean of the background outlier population was determined to be up to 0.8 magnitude fainter than the daily mean, with a standard deviation between 0.1 and 0.7 magnitudes. The quality of a given daily sequence is not known {\em a priori}, so the prior probability $Q$ that any given data point belongs to the foreground population is allowed to vary between 0.0 and 1.0.

Because the parameters are very close to being mutually orthogonal, we used an MCMC scheme employing a simple Metropolis-Hastings sampler. Two initial burn-in phases of $10^4$ samples were employed. At the end of each, the standard deviation of the parameter values was determined from the last $10^3$ steps to ensure better mixing in the next stage. A third production phase of $10^4$ steps yielded the posterior samples used to determine the daily extinction coefficients and to calculate the foreground membership probabilities for the individual data points. Median values of these samples are listed in machine-readable form in Appendix~\ref{sec:daily}, Table~\ref{tab:daily}. The mixture probabilities from Eq.~\ref{eq:marginal} provide clear thresholds against which the user can partition usable from unusable data, and are used as quality flags in  Appendix~\ref{sec:alldata}, Table~\ref{tab:alldata}.

\subsection{Differential extinction corrections}

The daily mean extinction coefficient defines the vertical extinction gradient across the diameter of the solar disk as a function of airmass. This gradient affects the measured solar radial velocity whenever the solar rotation axis is inclined relative to the vertical direction. This generally leads to a spurious redshift in the morning, when the approaching hemisphere of the Sun is lower in the sky than the receding hemisphere. The effect was first identified and corrected by \cite{1987ApJ...316..771D}, and has subsequently been modelled by the BiSON team \citep{2014MNRAS.441.3009D}. The analytic approach we adopt here was developed independently of, and is slightly different from, these previous studies, but the results obtained are essentially identical.

The difference in airmass $X$ between the upper and lower limbs of the Sun is
\begin{equation}
\delta X  = X_{\rm upper}-X_{\rm lower}
\end{equation}
where $X=\sec{z}[1-0.0012(\sec^2{z}-1)]$ \citep{1967AJ.....72..945Y}. The zenith angles $z_{\rm lower,upper}=z_\odot\pm \sin^{-1}(R_\odot/d_\odot)$ depend on the zenith angle $z_\odot$ of the solar centre, the solar radius $R_\odot$ and the Earth-Sun distance $d_\odot$ at the time of observation.

The corresponding difference in magnitudes of atmospheric extinction is $\delta m_{60}=k_{60}\delta X/2$ across the solar radius. We define a fractional difference in brightness $\epsilon\simeq\delta m_{60}\ln{10}/2.5$, with $\epsilon=0$ at the centre of the solar disc.  

The angle between the solar rotation pole and the local vertical is $\phi=PA-q$, where $PA$ is the position angle of the solar rotation axis relative to the north celestial pole at the time of observation. The parallactic angle $q$ is the angle subtended at the Sun by the celestial pole and the zenith, giving
\begin{eqnarray}
\sin{q} &=& \sin{H}\cos{\beta}/\sin{z_{\odot}}\\
\cos{q}&=&\sqrt{1-\sin^2{q}},
\end{eqnarray}
where $H$ is the solar hour angle and $\beta$ is the latitude of the TNG.

For a linear limb-darkening law with limb darkening coefficient $u$, the form of the rotation profile in the absence of differential extinction is \citep{1992oasp.book.....G}:
\begin{equation}
f(x)=\int^{\sqrt{1-x^2}}_{-\sqrt{1-x^2}} (1-u+u\mu(x,y))\ dy,
\end{equation}
where the brightness is $I(x,y)=1-u+u\mu(x,y)$ and the foreshortening cosine $\mu(x,y)\equiv\sqrt{1-x^2-y^2}$.

The orthogonal Cartesian coordinates $x$ and $y$ are in the directions of increasing solar rotation velocity and the solar north pole respectively in the plane of the sky. For solid-body rotation with equatorial velocity $v_{\rm eq}$, the brightness-weighted correction to the solar radial velocity is 
\begin{equation}
\delta v_{\rm r}=v_{\rm eq}\frac{\int^{1}_{-1}  x f(x) dx}{\int^{1}_{-1} f(x) dx}.
\label{eq:f}
\end{equation}
In the absence of differential extinction across the rotation profile, the integrand in the numerator of Eq.~\ref{eq:f} is an odd function, so the brightness-weighted velocity offset is zero.

We modify the calculation of the limb-darkened rotation profile to include such a density gradient, parameterised by the fractional extinction difference $\epsilon$ across the solar radius. A brightness gradient of uniform slope $\epsilon$ in $x$ and $y$ at an angle $\phi$ to the local vertical modifies the rotation profile in Eq.~\ref{eq:f} to a new form:
\begin{equation}
g(x,\phi)=\int^{\sqrt{1-x^2}}_{-\sqrt{1-x^2}} I(x,y)(1+\epsilon(-x\sin{\phi}+y\cos{\phi}))\ dy.
\end{equation}
The sign of the $x\sin{\phi}$ term should be negative if $\phi$ is positive west of the meridian.

The velocity bias $\delta v_{\rm r}$ introduced by the extinction gradient thus depends on $v_{\rm eq}$, $u$, $\phi$, and $\epsilon$ according to
\begin{eqnarray}
\delta v_{\rm r}&=&\frac{\int^{1}_{-1}  x g(x,\phi) dx}{\int^{1}_{-1} g(x,\phi) dx}v_{\rm eq}\nonumber\\
&=&\frac{(7u-15)\epsilon\sin{\phi}}{20(3-u)}v_{\rm eq}.
\label{eq:solidbody}
\end{eqnarray}

So far we have assumed that the solar photosphere rotates as a solid body, ignoring the effects of differential rotation, axial inclination and centre-to-limb variations in convective blueshift. \cite{2016A&A...595A..26R} modelled the effects of these modifications to the solar rotation profile on the solar radial velocity in the presence of an extinction gradient. Following their example, we can include differential rotation in our model by allowing the apparent rotational angular frequency $\omega_{\rm obs}$ to vary as a function of heliographic latitude $b$. This alters eq.~\ref{eq:f} such that the intrinsic solar equatorial rotation speed $v_{\rm eq}$ is replaced by $R_\odot\omega_{\rm rot}(x,y)$. We adopt the sidereal differential rotation law of \cite{1990ApJ...351..309S} as a function of heliographic latitude $b$, defining
\begin{equation}
\omega_{\rm rot}(\sin{b}) = A + B\sin^2{b} + C\sin^4{b},
\end{equation}
with 
$A=2.972\times 10^{-6}\ {\rm rad\ s}^{-1}$, 
$B=-0.484\times 10^{-6}\ {\rm rad\ s}^{-1}$ and 
$C=-0.361\times 10^{-6}\ {\rm rad\ s}^{-1}$. 
When the solar rotation axis is inclined at angle $i$ to the line of sight, a point at location $(x,y)$ on the visible solar hemisphere has
\begin{equation}
\sin{b}=y\sin{i}+\sqrt{1-x^2-y^2}\cos{i}.
\end{equation}
The integration for $v_{\rm rot}$ becomes
\begin{equation}
\delta v_{\rm rot}=\frac{\int^{1}_{-1}  x h(x,\phi) dx}{\int^{1}_{-1} g(x,\phi) dx}R_\odot\sin{i}
\label{eq:diffrot}
\end{equation}
where
\begin{equation}
h(x,\phi)=\int^{\sqrt{1-x^2}}_{-\sqrt{1-x^2}} I(x,y)\ \omega_{\rm rot}(x,y)\ (1+\epsilon(-x\sin{\phi}+y\cos{\phi}))\ dy.
\end{equation}

To take the Earth's instantaneous orbital motion and the centre-to-limb variation in convective radial velocity into account we must include additional terms:
\begin{equation}
\delta v_{\rm r}=\frac{\int^{1}_{-1}  (x h(x,\phi)R_\odot\sin i +c(x,\phi))dx}
{\int^{1}_{-1} g(x,\phi) dx}
-\frac{\int^{1}_{-1}  x g(x,\psi)dx}
{\int^{1}_{-1} g(x,\psi) dx}R_\odot\,\omega_{\rm orb}.
\label{eq:includeconv}
\end{equation}
Here $\psi$ is the position angle between ecliptic north and the local vertical at the Sun's location. It always differs from $\phi$ by less than the 7-degree tilt of the solar rotation axis relative to the ecliptic. The position angle of ecliptic north relative to celestial north at the Sun is given by 
\begin{equation}
    \sin{\rm PA_{ecl}}=\sin\varepsilon\cos\alpha
\end{equation}
where $\varepsilon$ is the 23.4-degree obliquity of the ecliptic and $\alpha$ is the solar RA at the time of observation. By analogy with the position angle of the solar rotation pole, $\psi = {\rm PA_{ecl}}-q$ where $q$ is the parallactic angle. The integration is specified separately in Eq.~\ref{eq:includeconv} in a Cartesian coordinate system in which $y$ is oriented toward the ecliptic north pole and $x$ is in the direction of Earth's orbital motion. The solution has the same form as Eq.~\ref{eq:solidbody} with $v_{\rm eq}$ replaced by $-R_\odot\,\omega_{\rm orb}$.

The convective contribution is
\begin{equation}
c(x,\phi)=\int^{\sqrt{1-x^2}}_{-\sqrt{1-x^2}} I(x,y)\ v_{\rm conv}(\mu(x,y))\ (1+\epsilon(-x\sin{\phi}+y\cos{\phi}))\ dy.
\end{equation}
Both $I(x,y)$ and $v_{\rm conv}(\mu(x,y))$ are azimuthally-symmetric functions of radius only. The contribution of $c(x,\phi)$ to $\delta v_{\rm r}$ in Eq.~\ref{eq:includeconv}\ is therefore independent of $\epsilon$ and $\phi$, being a constant that depends on the radial variations in limb darkening and convective velocity only. 
To illustrate this point by way of example, we approximate the centre-to-limb dependence of the line-of-sight convective velocity as a sigmoidal function with the form
\begin{equation}
v_{\rm conv}(x,y) = 0.4 -1.2\mu^2+0.6\mu^4\ {\rm km~s}^{-1},
\end{equation}
the coefficients being chosen to approximate Fig.~18 of \cite{2013A&A...558A..49B} and to  give
a derivative with respect to $\mu$ equal to zero at the centre and limb.
The double integral
\begin{equation}
\delta v_{\rm c} =\int^1_{-1}c(x,\phi)dx 
\end{equation}
has the analytic solution $\delta v_{\rm c}= 0.376991 - 0.273631 u$ for any linear extinction gradient across the solar disc. Since there is no dependence on either  $\epsilon$ or $\phi$, the extinction gradient across the convective velocity pattern does not produce any change in radial velocity with either airmass or position angle. We therefore treat the effect of the centre-to-limb variation in convective velocity as a time-invariant offset to the velocity zero-point, and evaluate the integrals in Eqs.~\ref{eq:diffrot} and \ref{eq:includeconv} to obtain
\begin{eqnarray}
\delta v_{\rm r}&=&\epsilon\sin{\phi}R_\odot\sin{i}\left(
\frac{ A(7u-15)}{20(3-u)}\right.\nonumber\\
&+&\frac{ B(19u-35+\cos^2{i}(3u-35))}{280(3-u)}\nonumber\\
&+&\left.\frac{C(187u-315-\cos^2{i}(630-118u)-105\cos^4{i}(u-1))}{6720(3-u)}
\right)\nonumber\\
&-&\frac{(7u-15)\epsilon\sin{\psi}}{20(3-u)}R_\odot\omega_{\rm orb}
\label{eq:drcorrect}
\end{eqnarray}

The velocity correction for differential extinction varies almost linearly with airmass, and has an amplitude of 2 to 3 m~s$^{-1}$ in an uninterrupted day's observations. As $\cos{i}$ varies through the year, the values of $\delta v_{\rm r}$ depart by only 1 or 2 mm s$^{-1}$ from the solid-body rotation values obtained using a sidereal rotation rate $\omega_{\rm sid}=2.875\times 10^{-6}\ {\rm rad\ s}^{-1}$. The inclination dependence involves only even powers of $\cos i$, and does not change sign when the Earth crosses the solar equator.

The radial-velocity plots in the lower-left panels of Figs.~\ref{fig:qualplot_clear}, \ref{fig:qualplot_calima} and \ref{fig:qualplot_cirrus} have all been corrected for this effect, using Eq.~\ref{eq:drcorrect} with $u=0.6$. The full set of extinction-corrected instrumental radial velocities are plotted in Fig.~\ref{fig:clpbary} for the duration of the observing campaign to date. 

\subsection{Intra-day solar radial-velocity variability}

\begin{figure}
\includegraphics[width=\columnwidth]{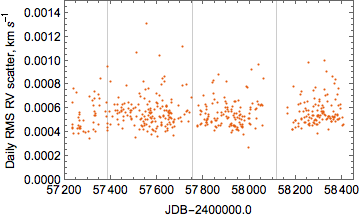}
\includegraphics[width=\columnwidth]{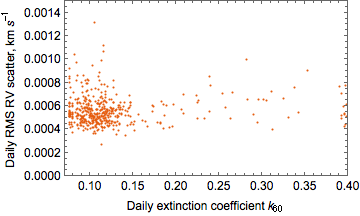}
\caption{Daily residual RMS radial-velocity scatter for all days with more than 48 observations (4 hours) with 99 percent or better mixture probabilities (as defined in Eq.~\ref{eq:marginal}). The same data are shown in the lower panel, plotted against the daily extinction coefficient.}
\label{fig:vnewrms}
\end{figure}

\cite{2016SPIE.9912E..6ZP} have shown that the 5-minute solar $p$-mode oscillations are clearly detected in sequences of 20-s exposures taken as part of the science verification programme for the HARPS-N solar telescope. Although the 5-minute exposures used in the longer-term study reported here are designed to suppress these oscillations, the daily radial-velocity traces in the lower-left panels of Figs.~\ref{fig:qualplot_clear} and \ref{fig:qualplot_calima} show that short-term correlated noise is still present. The daily RMS scatter shown in the upper panel of Fig.~\ref{fig:vnewrms} is nonetheless only slightly greater than the 0.43 m~s$^{-1}$ estimated uncertainty on a single radial-velocity measurement in conditions of good transparency. This agrees well with early results studied by \cite{2015ApJ...814L..21D}.

To ensure that the correlated noise is not caused by variable atmospheric transparency in {\em calima} conditions, in the lower panel of Fig.~\ref{fig:vnewrms} we plot the same values against the daily extinction coefficient. {\em Calima} is characterised by values of $k_{60}$ in the approximate range from 0.15 to 0.40 magnitudes per airmass. No dependence of the daily RMS scatter on extinction coefficient is apparent.

\begin{figure}
\includegraphics[width=\columnwidth]{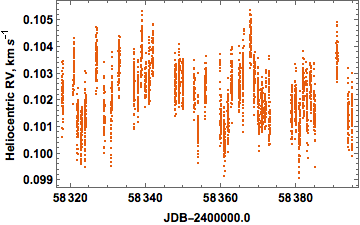}
\includegraphics[width=\columnwidth]{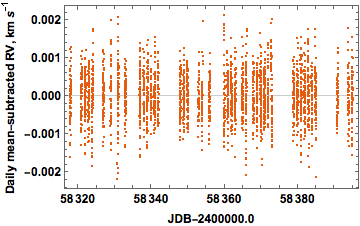}
\includegraphics[width=\columnwidth]{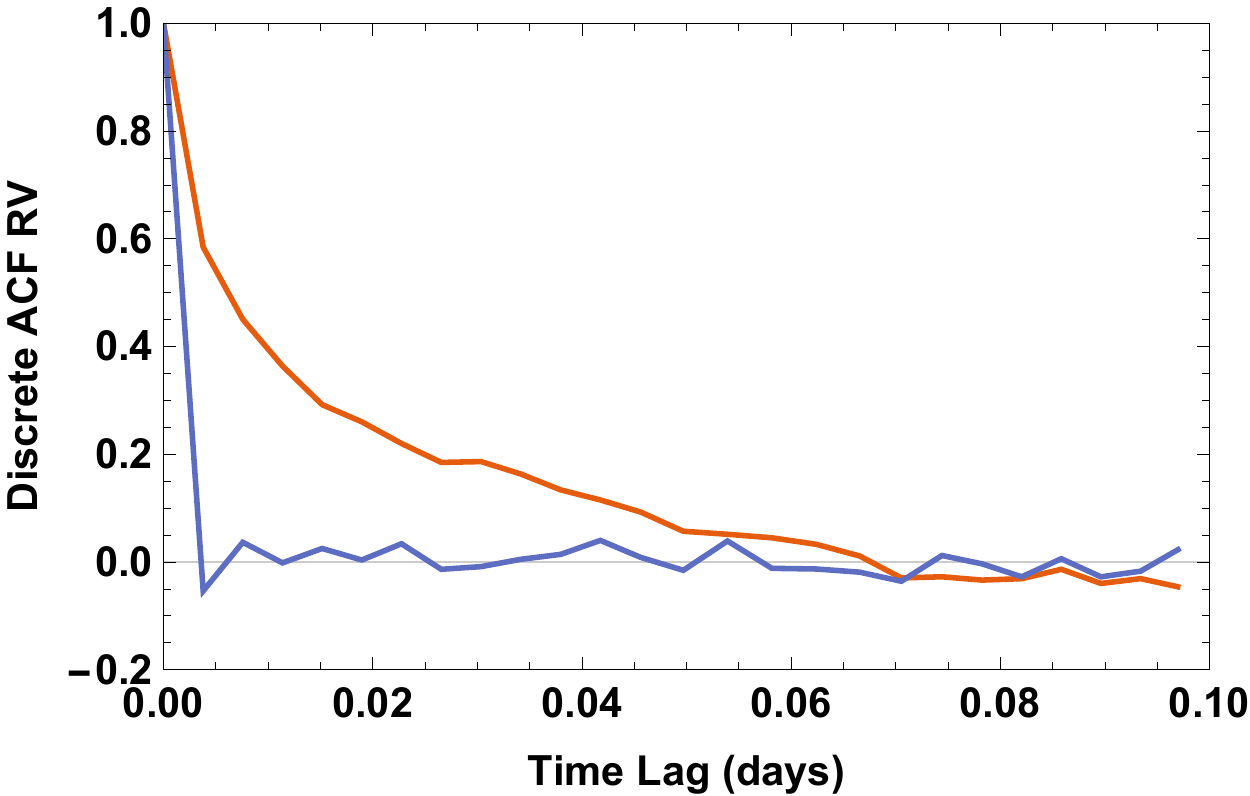}
\caption{Upper panel: heliocentric radial velocities for all days with more than 48 observations (4 hours) with 99 percent or better mixture probabilities, during  the period of low solar activity from 2018 Jul 21 to Oct 12. Blue arrows denote disc-centre passages of weak active regions. Middle panel: The same observations are shown with the daily mean values subtracted. Bottom panel: The Discrete Autocorrelation Functions of the mean-subtracted velocities (orange) and of white noise with the same time distribution and variance (blue) are shown.}
\label{fig:acfrvday}
\end{figure}

To obtain the autocorrelation function of the Sun on minutes-to-hours timescales in the quietest possible conditions, we selected the period from 2018 July 21-October 12. Daily SDO magnetograms show that six weak or decaying bipolar magnetic regions crossed the centre of the solar disc during this period. These disc-centre crossings are marked by blue arrows in the upper panel of Fig.~\ref{fig:acfrvday}. We estimated the RMS scatter of the individual 5-minute radial-velocity measurements under these quiet-Sun conditions by taking all days on which we obtained more than 4 hours' data (48 observations) with mixture probabilities (cf. Eq.~\ref{eq:marginal}) greater than 99 percent. Within each day we applied an iterative 4-sigma clip to remove very occasional outlying RV measurements with causes other than transparency fluctuations, which had not been rejected by the mixture modelling.

Even during periods when no discernible magnetic regions are present, the day-to-day scatter of the radial velocities greatly exceeds the intra-day scatter. The BIS and FWHM, however, display day-to-day scatter no greater than their daily averages.  HARPS-N observations entail comprehensive daily calibration sequences. These establish the geometry for the extraction of the spectral orders from the CCD frames, and the wavelength dispersion relation for the extracted spectra. The accuracy of the zero point of the radial velocities therefore depends on the accuracy of the daily calibration. Our results indicate that the zero-point accuracy of the daily calibrations is very similar to the 0.4 m~s$^{-1}$ RMS scatter 
measured by \cite{2018arXiv180901548D} in HARPS data reduced with the same version of the DRS used to extract and calibrate the solar spectra. We conclude that the daily jumps in the RV are likely to be a data-reduction artifact.

To eliminate the effects of day-to-day calibration zero-point errors and any residual rotationally-modulated stellar activity, we subtracted the daily mean from each day's RV measurements. The resulting set of velocities is plotted in the middle panel of Fig.~\ref{fig:acfrvday}. The autocorrelation function (ACF) of these data was computed using the method of \cite{1988ApJ...333..646E}, and is plotted in the lower panel of the same figure. For comparison, the ACF of white noise with the same variance, sampled at the same times, is also shown.

The ACF confirms clearly that the radial velocities are strongly correlated, with a correlation half-life close to 15 minutes. This is somewhat longer than the average lifetime of a solar granule. The radial velocities become effectively uncorrelated for time lags longer than about 0.07 day (1.7 hours). 

The 5-minute cadence of the observations is designed to suppress short-term variability caused by the solar 5-minute $p$-mode oscillations. For stellar observations, the HARPS and HARPS-N Rocky Planet-Search projects use 2 or 3 exposures of at least 15 minutes, spaced at least 2 or 3 hours apart. The 15-minute exposure mitigates the effects of $p$-mode oscillations, and the 2 to 3-hour sample spacing ensures that granulation noise is effectively uncorrelated \citep{2011A&A...525A.140D,2015A&A...583A.118M}. The autocorrelation properties of the HARPS-N solar data illustrate the need for such a strategy.

\begin{figure}
\includegraphics[width=\columnwidth]{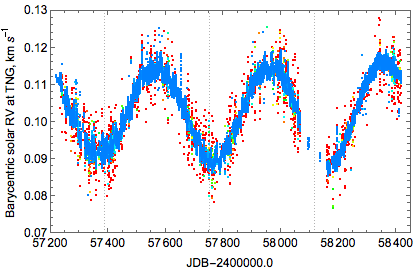}
\includegraphics[width=\columnwidth]{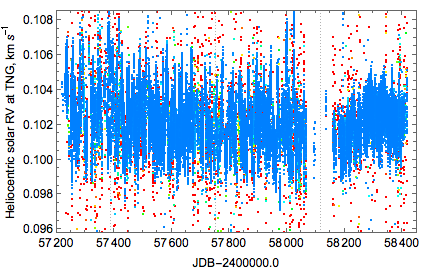}
\caption{Top: HARPS-N radial velocities in the barycentric reference frame. The solar reflex motion appears from Earth as a sinusoid with an amplitude of 12 m~s$^{-1}$ and a period close to 13 months. Bottom: HARPS-N radial velocities in the heliocentric reference frame. The solar reflex motion due to the planets has been removed. Vertical gridlines denote calendar-year boundaries at the start of 2016, 2017 and 2018. The points are colour coded for data quality as in Fig.~\ref{fig:qualplot_clear}, with foreground (good, blue) data overplotted on background (bad, red) data.}
\label{fig:clpbary}
\end{figure}

\section{Corrections for solar-system dynamics}
\label{sec:corrections}

\subsection{Solar barycentric motion}
\label{sec:barycentric}

The radial velocities derived from the solar spectra are processed using the same HARPS-N Data Reduction System that is employed for night-time observations. The exposure-weighted mid-time of each exposure is converted from UTC to Barycentric Julian Date in the Barycentric Dynamical Time standard, and the radial velocity is transformed to the reference frame of the solar-system barycentre. For both calculations, the apparent J2000 position of the centre of the Sun replaces conventional J2000 stellar coordinates.

When the target is the Sun, the use of the Barycentric Julian Date remains valid to within 2 or 3 seconds. The radial velocity, however, represents the component of the Sun's barycentric motion in the direction of the observer, in this case HARPS-N at the TNG. The raw radial velocities thus comprise the radial component of the total solar reflex motion induced by the planets. The sum of these motions is clearly seen in Fig.~\ref{fig:clpbary} as a sinusoidal variation with an amplitude of 12 m~s$^{-1}$ and a period close to 13 months. This is the synodic period of Jupiter observed from Earth.

To eliminate this reflex motion, we transform the radial velocities to the heliocentric system, using the JPL Horizons ephemeris \citep{1996DPS....28.2504G} to obtain the component of the Sun's barycentric motion in the direction of the TNG. This is subtracted from the raw velocities, reducing the observed velocities to a reference frame in which the rate of change of distance of the solar centre from the TNG is zero. The resulting radial velocities are shown in the lower panel of Fig.~\ref{fig:clpbary}. The average observed velocity of 102 m~s$^{-1}$ includes the instrumental zero point, the solar gravitational redshift and the granular convective blueshift, and agrees with the mean values determined by \cite{2013MNRAS.429L..79M} and \cite{2016MNRAS.457.3637H} using HARPS on the ESO 3.6-m telescope within 2 m~s$^{-1}$. 

%{\em Show plot of raw daily-average Barycentric DRS velocities. Describe use of JPL Horizons to model solar barycentric motion along sightline Sun-TNG.}

\begin{figure}
\includegraphics[width=\columnwidth]{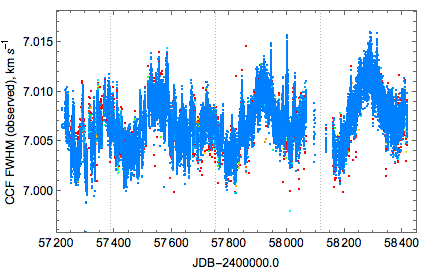}
\includegraphics[width=\columnwidth]{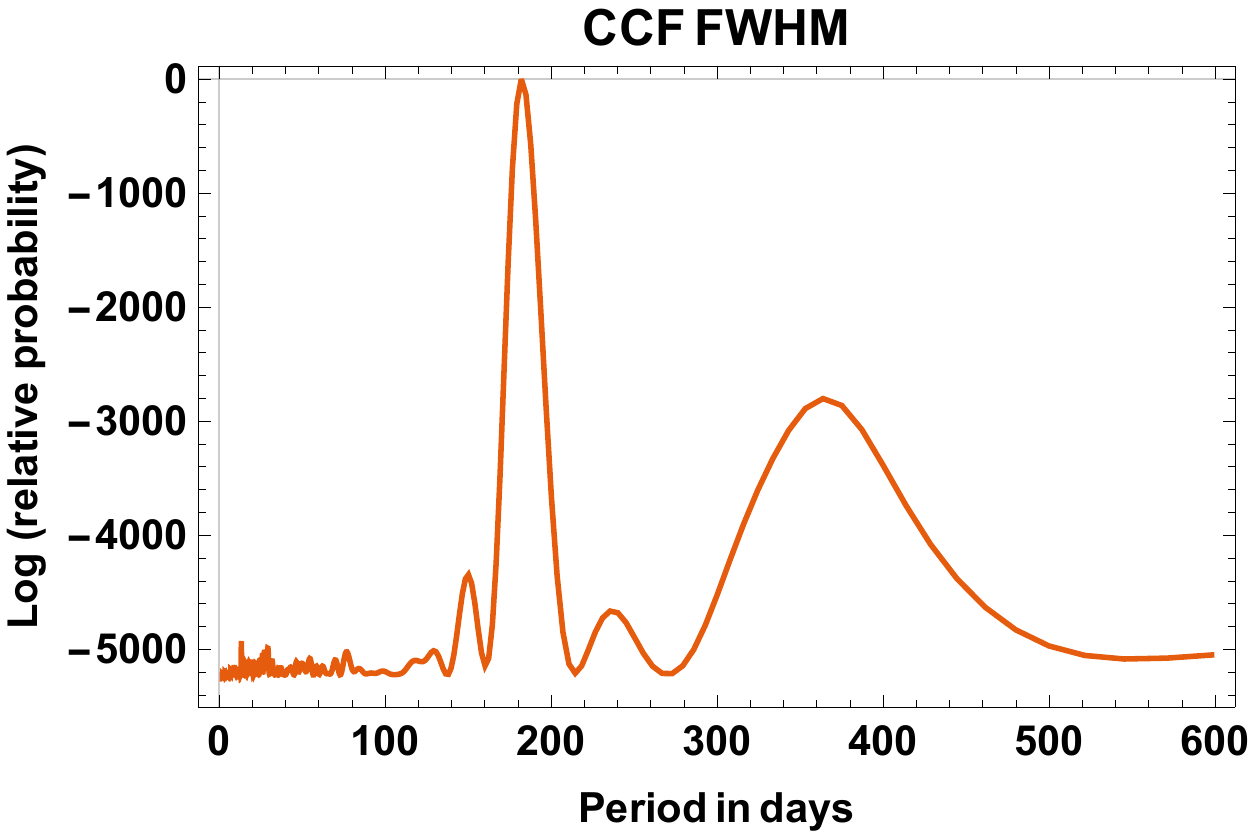}
\includegraphics[width=\columnwidth]{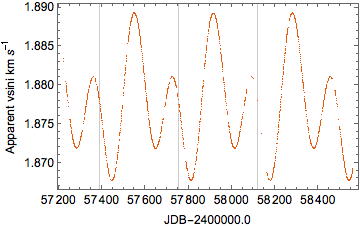}
\caption{The observed FWHM of the solar CCF (top panel) is dominated by an annual variation in the solar apparent angular velocity arising from the Earth's orbital eccentricity and a six-month oscillation in the projected equatorial velocity arising from the obliquity of the ecliptic plane relative to the solar equator. The middle and lower panels show the BGLS periodogram of the FWHM and the annual variations in the apparent solar $v\sin{i}$. Vertical gridlines denote calendar-year boundaries.}
\label{fig:clphwmf}
\end{figure}

\subsection{Earth orbital eccentricity and obliquity}
\label{sec:earthorbit}

The FWHM of the Gaussian fit to the CCF is known to be rotationally modulated in exoplanet host stars with moderate to high levels of stellar activity \citep{2009A&A...506..303Q}. The periodogram of the observed FWHM of the solar CCF, constructed with the Bayesian Generalised Lomb-Scargle (BGLS) formalism of \cite{2015A&A...573A.101M}, however, displays a clear additional modulation on timescales of 6 months and 1 year (Fig.~\ref{fig:clphwmf}).

These additional modulations are of dynamical origin. The  solar apparent equatorial velocity is $v\sin{i}=R_\odot\omega_{\rm obs}\sin{i}$, where $i$ is the inclination of the solar rotation axis to the line of sight. The  solar apparent angular velocity $\omega_{\rm obs}$ is the difference between the solar sidereal angular velocity $\omega_\odot$ and the instantaneous orbital angular velocity $\omega_\oplus(t)$ of the observer. The Earth's orbital eccentricity imposes an annual variation on the latter,
\begin{equation}
\omega_\oplus(t)=\frac{2\pi}{P_\oplus}\frac{a_\oplus^2}{r^2(t)}\sqrt{1-e^2}
\end{equation}
where $e$ is the Earth's orbital eccentricity, $P_\oplus$ is the  orbital period and $r(t)/a_\oplus$ is the Earth-Sun separation at time $t$ expressed in au.

\begin{figure}
\includegraphics[width=.9769\columnwidth, right]{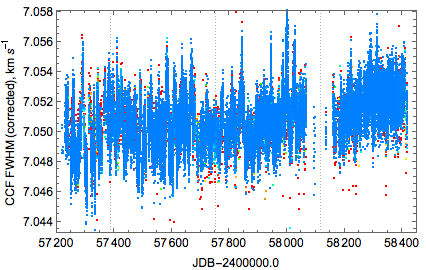}
\includegraphics[width=.9513\columnwidth, right]{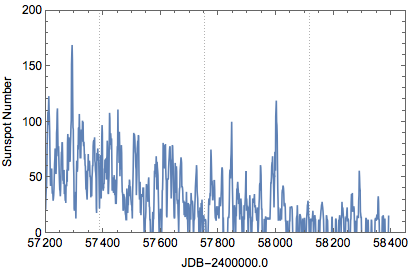}
\includegraphics[width=\columnwidth, right]{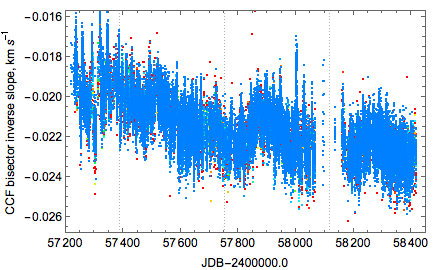}
\includegraphics[width=.9769\columnwidth, right]{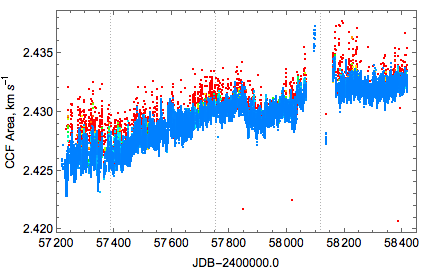}
\caption{Top to bottom: Time-domain variability in (a) the corrected FWHM in the sidereal frame; (b) the WDC-SILSO sunspot index; (c) the CCF bisector inverse slope and (d) the CCF area. The colour coding of the data points denotes data quality. Vertical gridlines denote calendar-year boundaries.}
\label{fig:timedomain}
\end{figure}

The variation in $v\sin i$ with time is straightforward to calculate given the Cartesian ecliptic direction vector $\hat{\boldmath{\omega}}_\odot$ of the solar rotation pole and the instantaneous Sun-Earth direction vector $\hat{\boldmath{r}}_{\oplus}$ in ecliptic coordinates, since $\sin i = \sqrt{1-(\hat{\boldmath{\omega}}_\odot . \hat{\boldmath{r}}_\oplus})^2$. The correction needed to recover the value of the FWHM that would be observed in the sidereal reference frame is, however, less straightforward. The observed CCF can be considered as the convolution of a limb-darkened rotation profile with the CCF that would be obtained for a hypothetical non-rotating star. For the purposes of this study we treat the observed FWHM $F_{\rm obs}$ as the quadrature sum of the intrinsic linewidth $F_0$ and an unknown fraction $\gamma$ of the observed $v\sin i$, such that
\begin{equation}
F^2_{\rm obs}\simeq F^2_0+ \gamma^2(\omega_\odot-\omega_\oplus)^2R_\odot^2\sin^2 i.
\end{equation}
An observer at interstellar distance in the Sun's equatorial plane would observe the sidereal FWHM
\begin{eqnarray}
F^2_{\rm sid}&\simeq& F^2_0+ \gamma^2 R_\odot^2 \omega_\odot^2\nonumber\\
&\simeq& F^2_{\rm obs}-\gamma^2 R_\odot^2 [(\omega_\odot-\omega_\oplus)^2\sin^2 i-\omega_\odot^2].
\end{eqnarray}
We find that the RMS scatter in the resulting values of $F_{\rm sid}$ is minimised for $\gamma = 1.04$. Using this value we plot the corrected FWHM in the sidereal frame in Fig.~\ref{fig:timedomain}. To first order, the CCF profile is the convolution of a local line profile with a stellar rotation profile of unit area, re-normalised to the resulting continuum level. So long as the area of the local line profile does not change strongly between facular and quiet-Sun areas, and the centre-to-limb variation in the area of the local line profile is small, the area $W$ of the CCF will remain approximately constant even when parts of the stellar disc are occupied by dark spots or bright faculae. The area of the fitted profile is thus proportional to the product of the FWHM and the contrast,
so $C_{\rm sid}F_{\rm sid}=C_{\rm obs}F_{\rm obs}$ gives the corresponding correction to the contrast. 

%{\em Describe corrections for to FWHM for Earth orbit eccentricity and solar obliquity.}

\section{Line-profile behaviour on the approach to solar minimum}
\label{sec:trends}

When corrected to the sidereal reference frame, the FWHM of the solar CCF shows a barely significant increase over the duration of the study. \cite{2018arXiv180901548D} found a similar slow change in the FWHM for HARPS, which they attributed to long-term changes in the focus of the instrument. The FWHM shows strong short-term variability, however, and peaks strongly at the times when one or more major spot groups are near the centre of the disk (Fig.~\ref{fig:timedomain}, top two panels; \citealt{2018SILSO,2016SoPh..291.2733C}). 

The Bisector Inverse Slope, plotted in the third panel of Fig.~\ref{fig:timedomain}, exhibits a secular decline over the three years of observation amassed to date, with a shorter modulation clearly seen at the solar rotation period and its first harmonic. Although it is likely that the BIS will be affected to some extent by the modulation of the solar $v\sin i$, Fig.~{\ref{fig:timedomain} shows that it is not the dominant signal.

This long-term trend is mirrored in the area of the Gaussian fit to the CCF, plotted in the bottom panel of Fig.~\ref{fig:timedomain}. 

\subsection{Long-term trends}

Early in the sequence the BIS shows rapid oscillations, as might be expected if multiple active regions cross the disc during every solar rotation. Underlying the peaks associated with disc-centre passages of large magnetic regions, the BIS exhibits a long-term downward trend. This trend appears to mirror the general decline in sunspot number over the same period. 

The CCF area shows a corresponding secular increase over the three years of observation amassed to date.
 This increase is not seen in the FWHM, so cannot be attributed to, for example, a long-term change in spectrograph focus. On shorter timescales of order weeks to months, the FWHM is anticorrelated with the CCF contrast. The CCF area, being the product of the contrast and the FWHM, does not therefore contain a strong rotational signal. The secular increase in the CCF area thus appears to be driven by a slow increase in the CCF contrast. This might happen if, for instance, the relative contributions of the blue and red echelle orders were changing slowly with time, as might happen through uncorrected ageing of a calibration lamp's colour temperature. We can rule out this possibility because the HARPS-N data-reduction system corrects the blaze function and normalises the continuum level in all orders before computing the CCF, with the explicit purpose of preventing such chromatic imbalances.

The trends in both the area and the BIS are interrupted temporarily at approximately BJD 2457850.0, and resume from their new levels from about BJD 2457900.0 (Fig.~\ref{fig:timedomain}). This interruption coincides with the appearance of the large, persistent bipolar active regions mentioned above (AR12653 and AR12654 in late 2017 March, then AR12674 and AR12675 early in 2017 September), following a lengthy period of low activity. From late 2017 onward, the decline toward solar minimum resumed, with few spot groups visible. 

At face value, a secular increase in the equivalent widths of the thousands of spectral lines that contribute to the CCF area gives the appearance of a global decrease in the stellar effective temperature. The area of the CCF represents a reliable proxy of the temperature and metallicity of a star, as demonstrated by \cite{2017MNRAS.469.3965M}. Under the reasonable assumption that the Sun did not change its metallicity during the three years spanning our observations, we can use the CCF area to track global temperature changes in the photosphere of the Sun. The correlation with the BIS suggests that the trends are also associated with a change in line asymmetry, and hence with changes in either the magnetic field \citep{1990A&A...231..221B,2017A&A...597A..52M} or the flux effect due to dark spots \citep{2010A&A...512A..39M,2016MNRAS.457.3637H}. 

Unlike the BIS, however, the CCF area shows little evidence of rotational modulation. This lack of rotational modulation suggests that the structures responsible for the change in CCF area are axisymmetrically distributed over the solar surface. We speculate that the CCF area may be tracing the evolution of the solar magnetic network, which largely comprises dispersed magnetic flux elements from decayed active regions being swept up towards the poles by meridional circulation. As the activity cycle declines, the supply of flux elements to the network decreases, reducing localised facular brightening and giving the appearance of a decreased effective temperature.

A decline in network flux also increases the dominance of the asymmetric quiet-Sun pattern of granular line asymmetry (e.g. \citealt{2013ApJ...763...95C}). The changes in line asymmetry measured by the BIS show both the long-term trend seen in the CCF area and a rotationally-modulated component that responds to the visibility of the facular magnetic fields of large bipolar regions. 

The connection between the effective temperature and the coverage of activity regions and network is not yet fully explored, but the solar data already offer some useful insights. The visibility of the more strongly RV-suppressive facular magnetic-field concentrations in SDO images has been found to be a stronger predictor of the solar radial velocity than the total visible-hemisphere coverage of active-region and network field elements \citep{2019arXiv190204184M}. \cite{2019arXiv190204184M} find the filling factor of the small, weakly RV-suppressive magnetic areas that they label ``network'' to be greater (and around solar minimum, much greater) than the filling factor of the more strongly RV-suppressive faculae. If the CCF area is sensitive to the temperature contrast  and filling factor of network/faculae in the optical, changes in the CCF area will mostly be driven by the slow, small, long-term variation of network rather than the much less dominant (by area) faculae. The BIS, on the other hand is sensitive to velocity suppression, so it would see both the long-term changes (due to changes in network and facular area) and the short term effects of rotationally modulated faculae.

\begin{figure}
\includegraphics[width=.85\columnwidth]{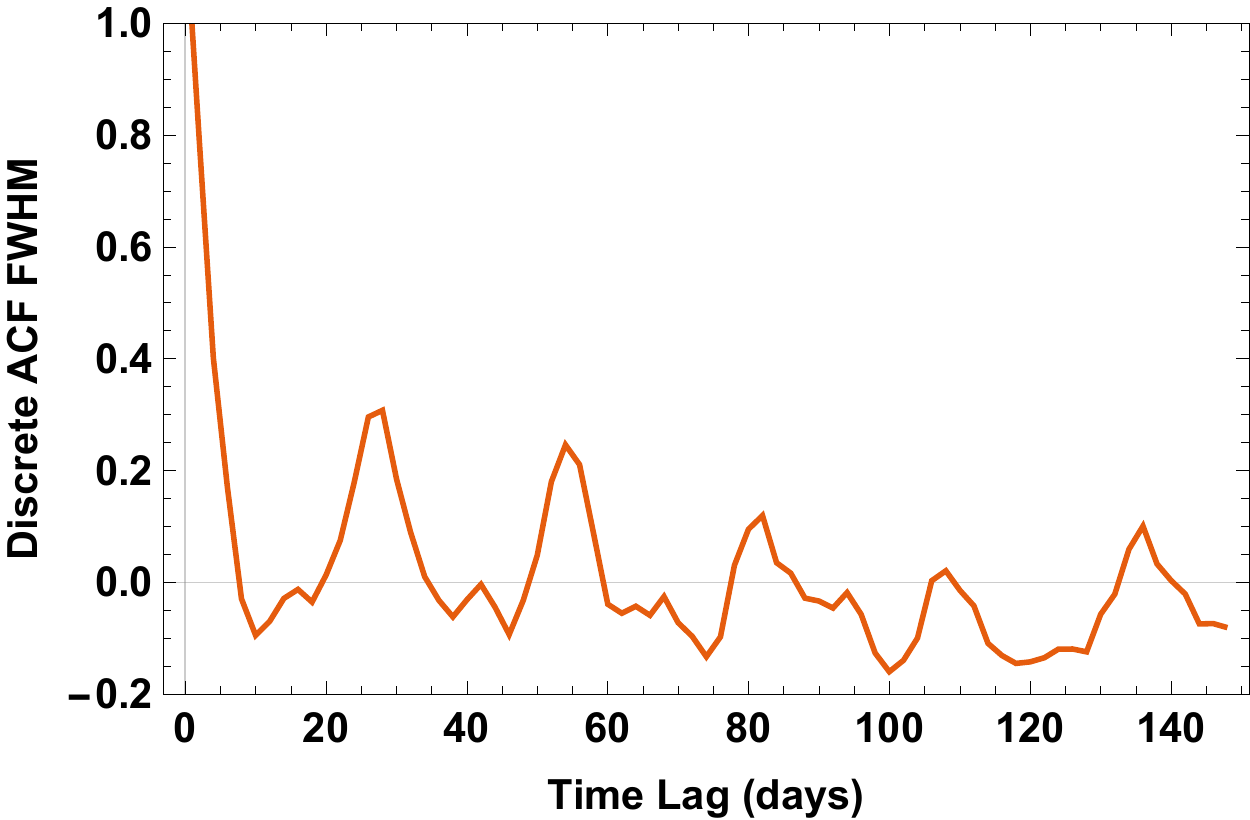}
\includegraphics[width=.85\columnwidth]{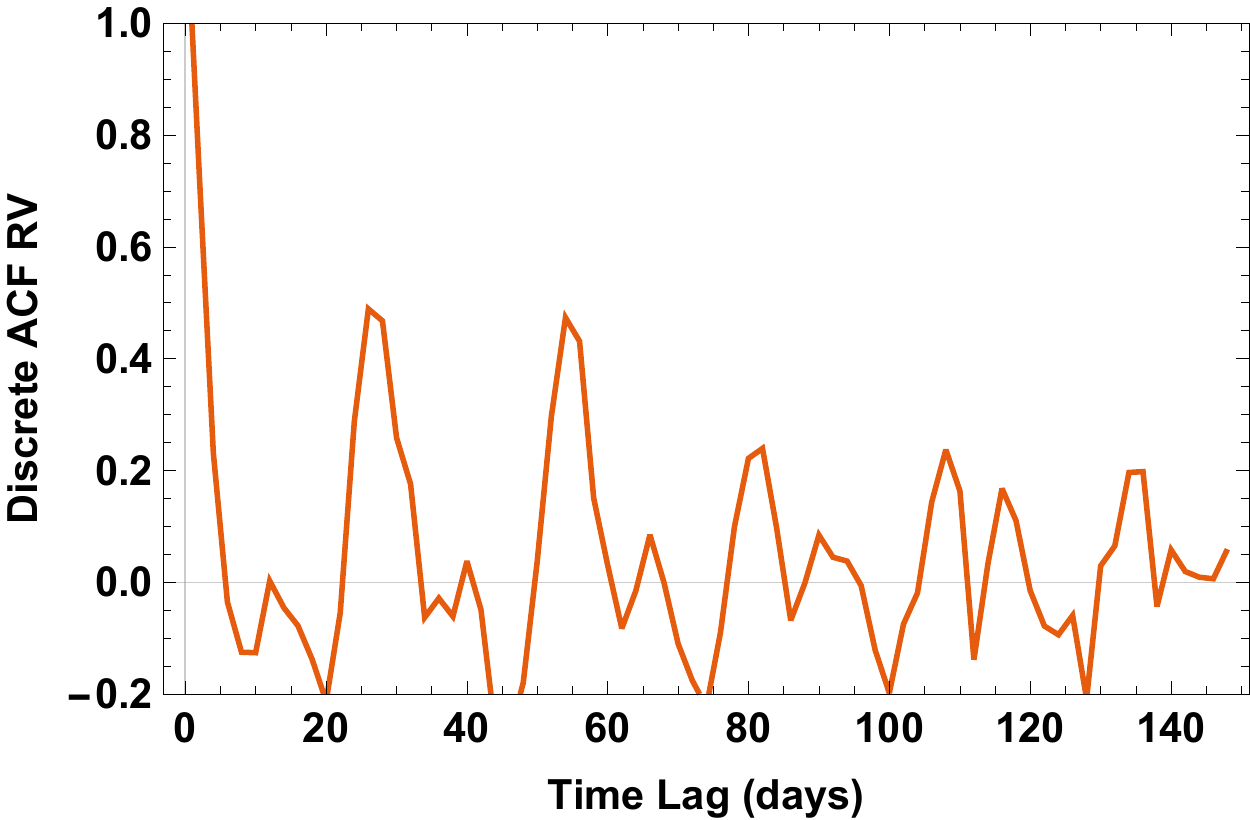}
\includegraphics[width=.85\columnwidth]{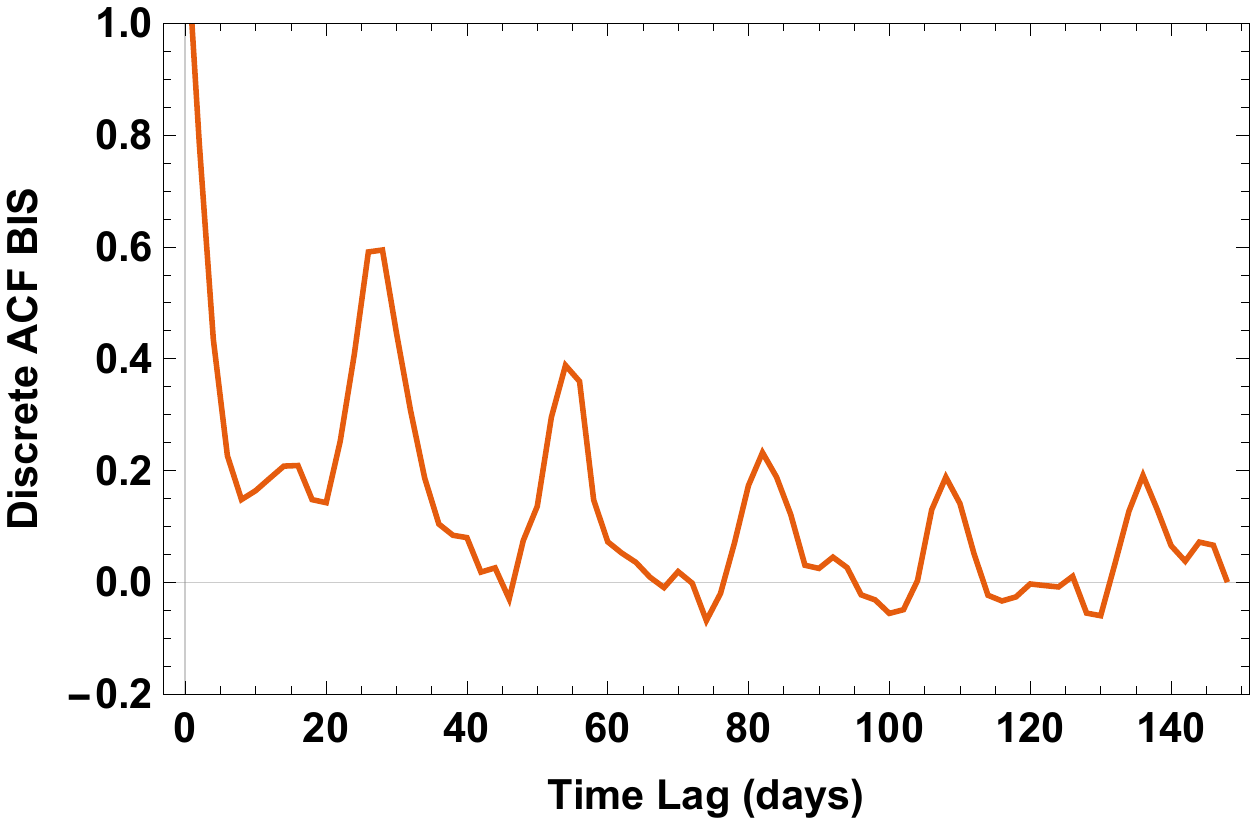}\\
\includegraphics[width=.85\columnwidth]{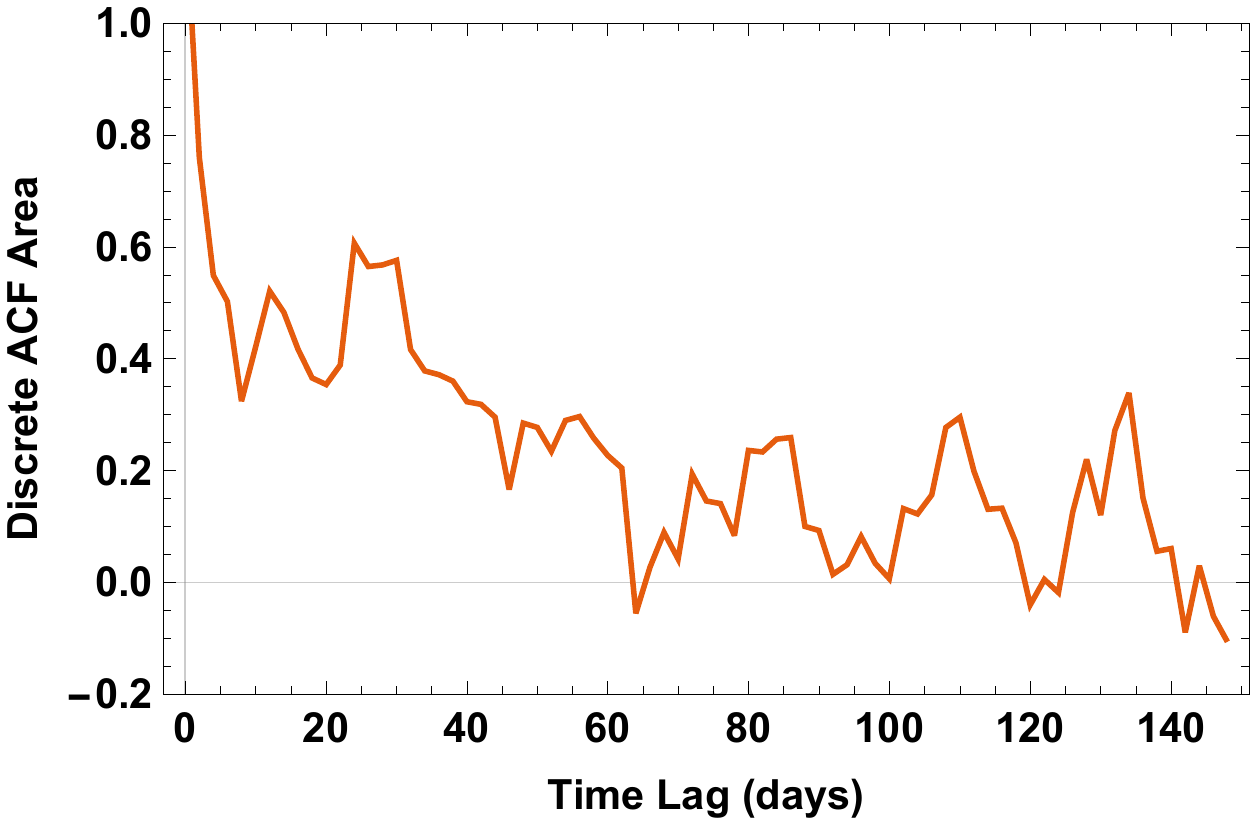}
\caption{Autocorrelation functions for (top to bottom) the FWHM, daily median radial velocity, BIS, and area of the CCF.}
\label{fig:autocorrelation}
\end{figure}

\subsection{Short-term line-profile variability}

The FWHM and the BIS both peak strongly at times when major sunspot groups and their accompanying bipolar magnetic regions cross the solar disc for the first time. Unlike the FWHM, however, the BIS shows a persistent periodic signal at the solar rotation period for about 3 rotations following the first passage of each major new spot group. 

\begin{figure}
\includegraphics[width=\columnwidth]{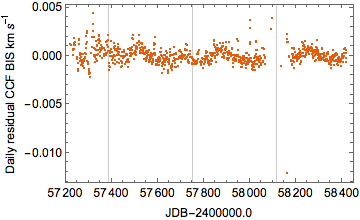}
\includegraphics[width=\columnwidth]{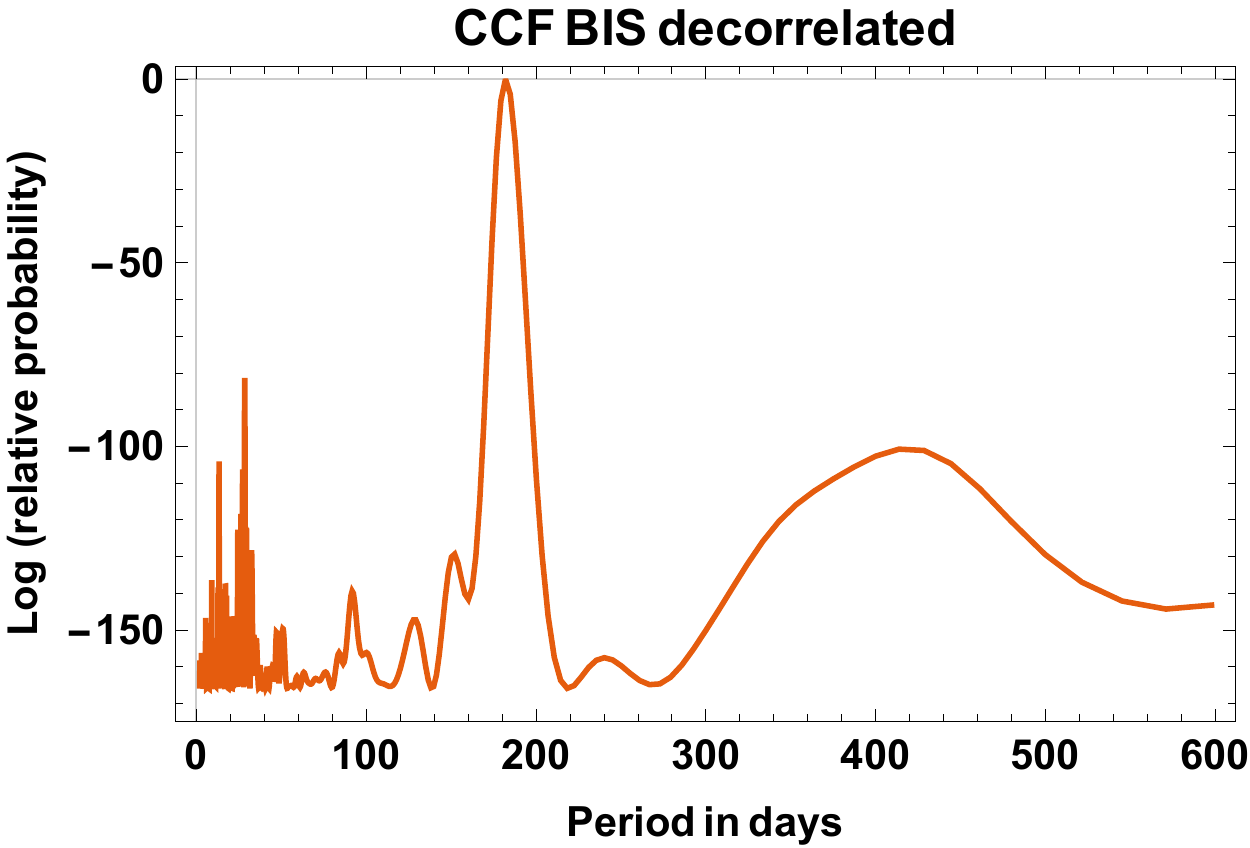}
\includegraphics[width=\columnwidth]{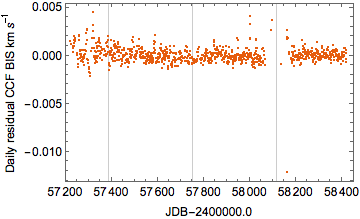}
\caption{Residual daily median values (top panel) of the BIS of the CCF after subtraction of the scaled long-term variation in the CCF area. The middle panel shows the Bayesian Generalised Lomb-Scargle periodogram of the residuals, showing strong power near 180 days and 1 year. The lower panel shows the residual BIS after decorrelation against CCF area and a pair of sinusoids with periods of six months and one year. Vertical gridlines denote calendar-year boundaries.}
\label{fig:bisfres}
\end{figure}

This behaviour is particularly apparent following the appearance of the pair of large bipolar solar active regions AR12654 and AR12655, which crossed the solar disc centre near JD 2457845.5 (2017 March 30). One rotation later, these regions had decayed to the extent that only remnants of the leading spots are visible in continuum images from SDO/HMI \footnote{https://www.solarmonitor.org/}. The SDO/HMI magnetograms, however, show that the bipolar fields from these active regions persisted for a further 3 or 4 rotations.

\begin{figure}
\includegraphics[width=\columnwidth]{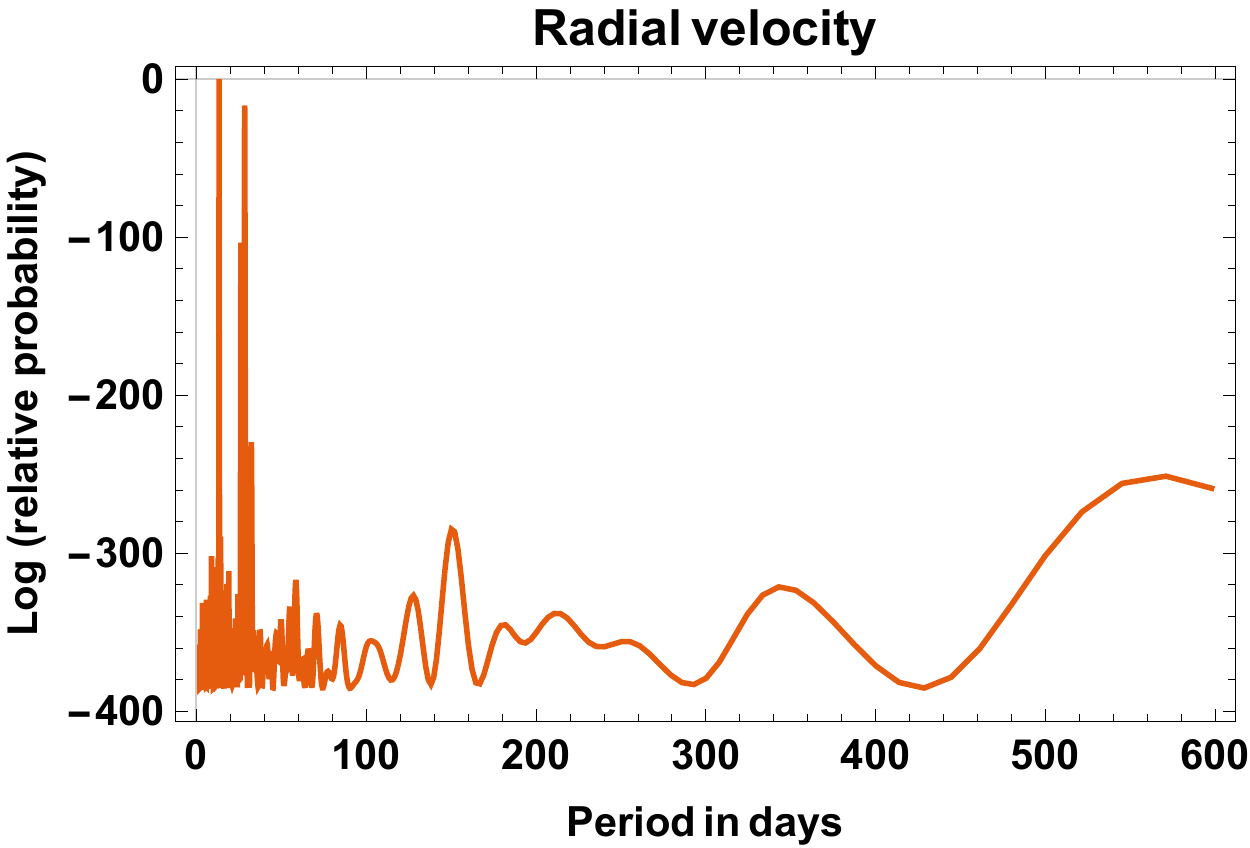}
\includegraphics[width=\columnwidth]{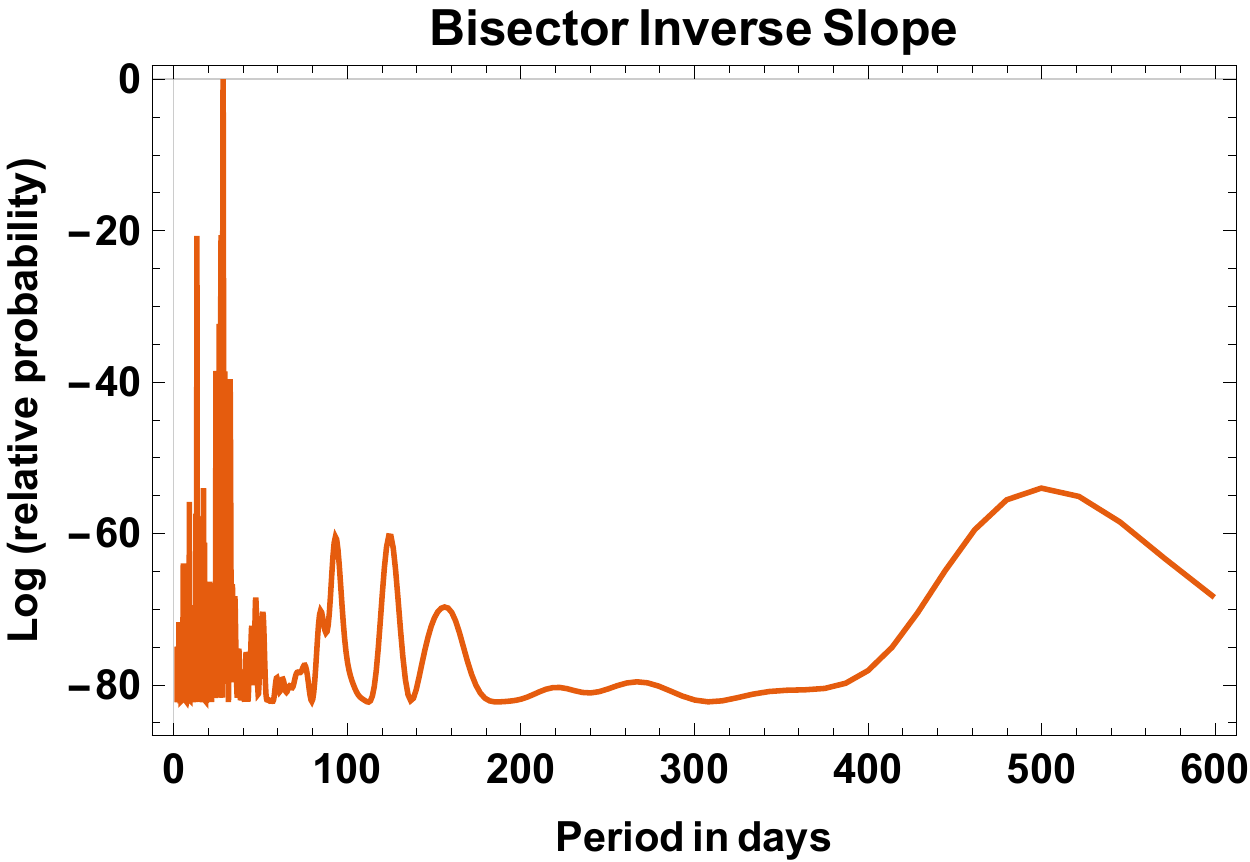}
\caption{Bayesian Generalised Lomb-Scargle periodograms for the daily median radial velocity (top), and bisector inverse slope (bottom) of the CCF.}
\label{fig:bgls}
\end{figure}

The same behaviour is seen following disc centre passage of the large bipolar regions AR12674 and AR12675 near JD 2458002.5 (2017 September 5). These same two spot groups survived to cross disc centre one Carrington rotation later on 2017 October 1, when they produced secondary peaks in both the FWHM and the BIS. A third ``echo" of this event appears strongly in the BIS plot (Fig.~\ref{fig:timedomain}) after a further Carrington rotation, even though the spots themselves had almost disappeared. SDO magnetograms taken on 2017 October 28 show that the bipolar magnetic regions associated with these groups are still very much present.

The coincidence of strong, isolated peaks in the FWHM time series with similarly isolated peaks in the sunspot number, and the greater observed persistence
of the BIS signal for several rotations after the emergence of each new active region suggests strongly that the FWHM responds to dark spots while the line asymmetry measured by the BIS traces inhibition of convection in active-region faculae.
To place this observation on a more quantitative footing we computed the autocorrelation functions of the timeseries data in RV, FWHM, BIS and CCF area. For each time series, we took the daily median value and time of observation, and subtracted a linear trend in time. We used the method of \cite{1988ApJ...333..646E} to obtain the ACF of the resulting irregularly-sampled timeseries, with a bin width of 2 days. 

The results are presented in Fig.~\ref{fig:autocorrelation}, for time lags from 0 to 150 days. The BIS shows the most persistent periodicity, with peaks occurring at multiples of 27 days, out to 110 days (4 solar rotations). The RV and FWHM show clear recurrence peaks out to 55 days (2 solar rotations). The autocorrelation function of the CCF area shows a steady decline with lag. There is little evidence of periodicity.

\subsection{Active-region BIS signal and radial velocity}

As discussed above, the BIS shows a secular trend mirroring  the smooth long-term change in the CCF area. To isolate the rotational signal in the BIS we decorrelated the BIS against the CCF area. The residual BIS signal exhibits annual and six-month periodicities similar to those seen in the FWHM (Fig.~\ref{fig:bisfres}). We fitted out these periodicities using sinusoids with periods of 6 months and 1 year, whose phases and amplitudes are to be treated as free parameters. The corrected BIS signal is shown in the lower panel of Fig.~\ref{fig:bisfres}.

%The BIS and the EW are seen to be strongly anticorrelated in the top panel of Fig.~\ref{fig:correlations}. 

%\begin{figure}
%\includegraphics[width=\columnwidth]{biseqwf.png}
%\includegraphics[width=\columnwidth]{bisvnew.png}
%\includegraphics[width=\columnwidth]{eqwvnew.png}
%\caption{Top: The equivalent width of the HARPS-N solar CCF is strongly anticorrelated with the bisector inverse slope over the duration of the study. Middle and bottom: the heliocentric radial velocity, corrected for differential extinction, shows weak correlations with BIS and EW respectively. The correlation of RV with BIS is smoother than that with EW.}
%\label{fig:correlations}
%\end{figure}

%{\em Discuss secular evolution of CCF bisector slope and equivalent width. Discuss and eliminate possible instrumental causes. Compare with classical  activity indicators (Ca II HK, sunspot number, 10.7-cm radio flux. What caused the “kink” starting JD2457850, peaking JD2457900? Why is bisector slope rotationally modulated but EW less so, even though they are anticorrelated?}

Periodograms of the daily median RV and de-trended BIS time series bear a close resemblance to each other (Fig.~\ref{fig:bgls}). Both periodograms are dominated by the solar rotation period and its first harmonic. 

%{\em Show and discuss periodograms and autocorrelation functions in RV, Bisector slope, FWHM and contrast. Note persistence of modulation signal over several solar rotations, indicating origins in faculae rather than short-lived spots.}

\begin{figure}
\includegraphics[width=\columnwidth]{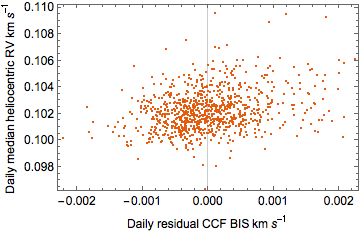}
\caption{Daily median radial velocity versus bisector inverse slope.}
\label{fig:bscovnew}
\end{figure}

\begin{figure}
\includegraphics[width=.9472\columnwidth, right]{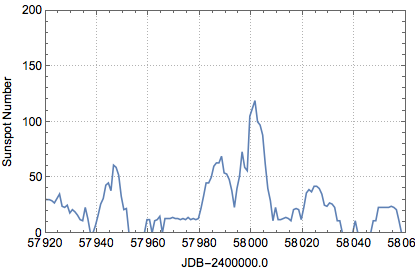}
\includegraphics[width=.9790\columnwidth, right]{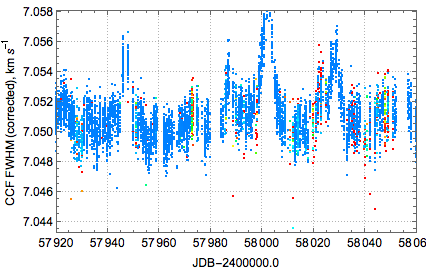}
\includegraphics[width=.9790\columnwidth, right]{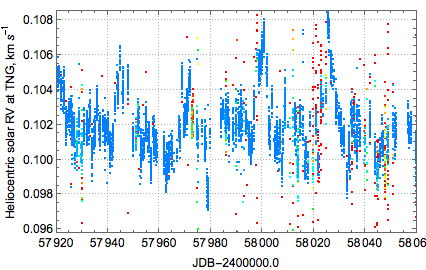}
\includegraphics[width=\columnwidth, right]{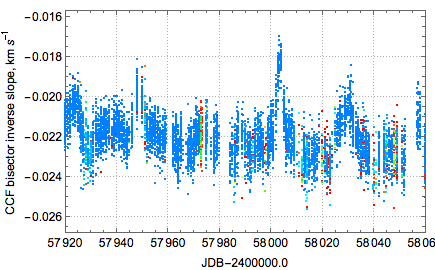}
\caption{Time-domain variability from 2017 mid-June to early November. Top to bottom: (a) the WDC-SILSO sunspot index; (b) the corrected FWHM in the sidereal frame; (c) heliocentric radial velocity and (d) the CCF BIS. The RV signatures of active-region typically peak $\sim 1$ and $\sim 3$ days before the FWHM and BIS signatures respectively. Colour coding denotes data quality.}
\label{fig:timezoom}
\end{figure}

\subsection{RV proxies and time lags}
\label{sec:lags}

Given the close resemblance of the RV and BIS periodograms, the similarity of the patterns of peaks in the RV and BIS in fig.~\ref{fig:timedomain} and the evidence described above that the oscillatory part of the BIS tracks the active-region facular filling factor, it is tempting to infer that the BIS will be a good proxy indicator of the radial velocity. Fig.~\ref{fig:bscovnew}, however, shows that the two are almost entirely uncorrelated.

This lack of correlation arises from the temporal behaviour of the RV and the line-profile parameters. Fig.~\ref{fig:timezoom} shows the sunspot number, corrected FWHM, radial velocity and uncorrected BIS during the period between 16 June and 3 November 2017, when a number of large, relatively isolated active regions crossed an otherwise quiet Sun. Close inspection of this figure, and of the discrete cross-correlation functions \citep{1988ApJ...333..646E} in Fig.~\ref{fig:dcflags}, reveals that the maxima in radial velocity occur 1 to 3 days before the maxima in FWHM and BIS respectively. Given that each peak has a width of about 7 days, this temporal offset between the RV and BIS weakens the correlation.

Similar temporal shifts between maxima in the RV, BIS, FWHM and chromospheric Ca II {\em H}\&{\em K} emission flux have been reported for other intensively-observed stars such as GJ674 \citep{2007A&A...474..293B}, GJ176 \citep{2009A&A...493..645F}, CoRoT-7 \citep{2009A&A...506..303Q} and HD41428 \citep{2014A&A...566A..35S}. While the authors of these papers generally attribute the phase lags in these cases to the RV signal being dominated by dark spots, the HARPS-N solar data indicate that the same phenomenon is present even when the RV signal is dominated by facular suppression of the convective blueshift \citep{2010A&A...512A..39M,2016MNRAS.457.3637H}. It is also worth noting that HD41248 is an inactive G2V star, so the observation by \cite{2014A&A...566A..35S} that increases in RV precede increases in FWHM by 2-3 days is consistent with the solar behaviour, i.e., facular, not spot, dominated  RV variations.

The reasons for this temporal offset between the RV and the line-profile parameters are subtle but important. The FWHM is expected to increase when a dark sunspot group crosses the centre of the solar disc. This phenomenon is well known from Doppler imaging of starspots \citep{1983PASP...95..565V}: when the missing light from a dark spot crosses the stellar rotation profile, it reduces the depth of the centre of the normalised line profile while approximately preserving the area (subject to any centre-to-limb variation in the area of the local CCF). The width of the line must therefore increase to compensate. The lack of persistence seen in the FWHM in Figs.~\ref{fig:timedomain} and \ref{fig:autocorrelation} supports an association of the FWHM with dark spots rather than the longer-lived surrounding magnetic regions. 

\begin{figure}
\includegraphics[width=\columnwidth]{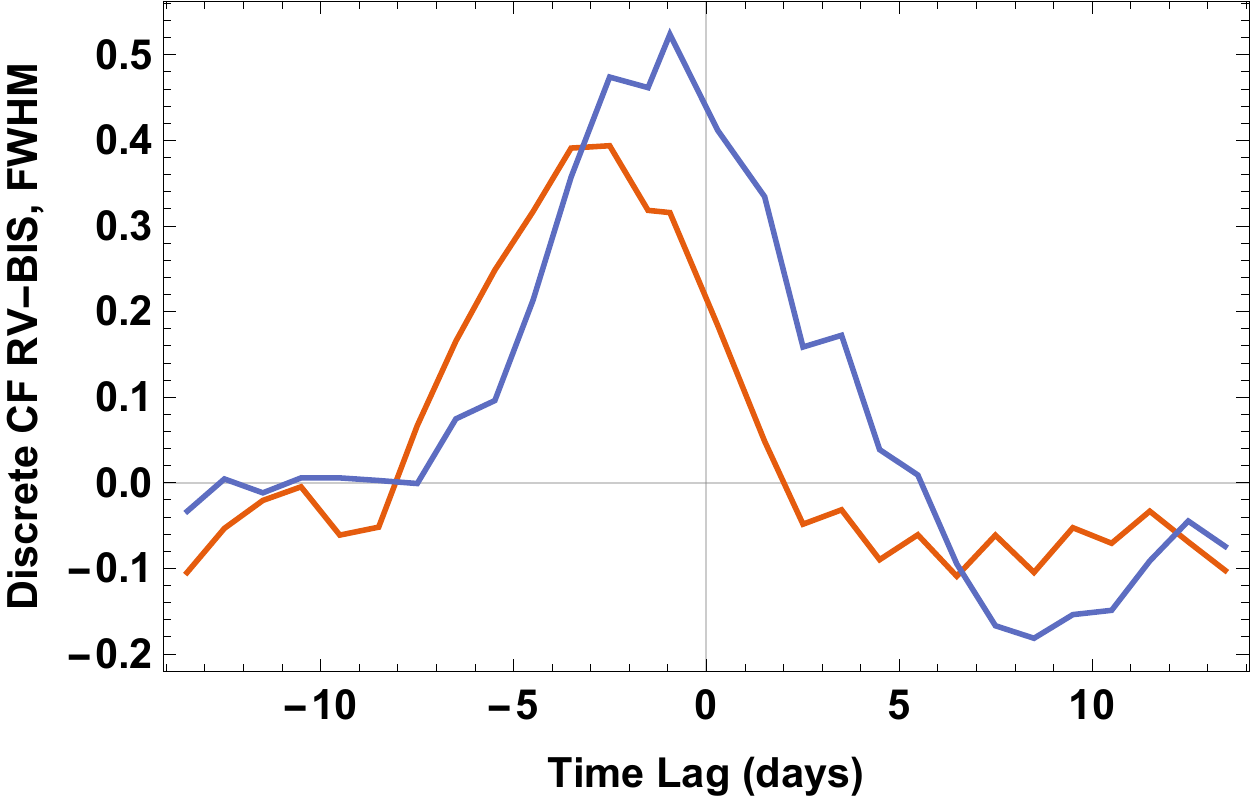}
\caption{Discrete time-series cross-correlation functions of RV against the corrected BIS (as in the lower panel of Fig.~\ref{fig:bisfres}, orange) and the sidereal FWHM (blue), for the full dataset of daily median values. The bin size is 1 day. The RV variations lead the BIS by 3 days, and the FWHM by 1 day.}
\label{fig:dcflags}
\end{figure}

In the absence of granular convection, we might expect the BIS to track the first derivative of the FWHM as the spectral signatures of dark spots cross the line profile from blue to red. Such an effect may be present at a low level, but in reality the BIS peaks only slightly later than the FWHM when an active region crosses disc centre. As noted above, the BIS peaks repeat for 3 or 4 solar rotations after the initial appearance of a large, isolated active region, even after the spots themselves have dispersed. This supports an association of the line asymmetry (quantified by the BIS) with the suppression of granular convection in the strongly-magnetic, bright facular regions that are the longer-lived counterparts of bipolar sunspot groups.

The close connection proposed by \cite{2010A&A...512A..39M} between the solar radial velocity and magnetic suppression of granular convection has been confirmed observationally by \cite{2016MNRAS.457.3637H} and \cite{2019arXiv190204184M}. The temporal offset between the RV and BIS signatures thus appears all the more surprising, until we consider how the measurements are actually made.

The cause of these temporal offsets was discussed by \cite{2014ApJ...796..132D} and is illustrated in their Figure 6. The HARPS-N DRS determines radial velocities by fitting a Gaussian to the CCF and recording the location of the peak. The CCF profile represents the convolution of many different local line profiles with the limb-darkened stellar rotation profile. As discussed above, a dark spot crossing the disc creates a disturbance to the CCF that is symmetric in time about disc-centre crossing  if the CCF profile itself is reasonably symmetric. If we now replace the dark spot with a highly-simplified active-region CCF which contributes similar total flux but whose line profile is displaced redward by suppression of convective blueshift, the net effect is to add a travelling perturbation whose form resembles the first derivative of the local line profile. Being antisymmetric, this disturbance influences the Gaussian fit, and hence the measured radial velocity, in a way that is not symmetric about disc-centre passage. 

For a slowly-rotating star like the Sun, the rotation profile is narrower than the intrinsic linewidth. When the active region is on the approaching hemisphere, the disturbance weakens the blue wing of the CCF and moves the core redward. When the active region is at disc centre, the core of the line moves redward. When on the receding hemisphere, the disturbance boosts the red wing of the profile and weakens the blue side of the core. The displacement of the centre of the fitted Gaussian is therefore always redward, but is not necessarily symmetric about the time of disc-centre passage. The time lags between the maxima of the RV, the BIS and the FWHM thus depend crucially on the details of changes to the local line asymmetry that occur as the active region crosses the disc.

There is another effect that could introduce time delays between different activity indicators. Strong, mostly vertical (buoyant) magnetic fields inhibit, in particular, flows
perpendicular to them.  These however, are not visible at disc centre. Their area and limb darkened visiblity peaks $\sim 40$ degrees off disc centre. This leads to a natural phase offset between BIS (which is more affected by magnetic inhibition of velocities, peaking away from from disk center, 
and suppressed relative to local quiet Sun) and FWHM (which is more affected by thermal variations, peaking at disc centre). One implication is that the lag would then be, at least in part,
a function of the rotation period.

Figure ~\ref{fig:timezoom} also shows that at times when the Sun is quiet, the day-to-day scatter in the FWHM and BIS is comparable to the intra-day scatter. The same is not true of the radial velocity, which shows an additional day-to-day scatter of order 0.4 m~s$^{-1}$. The most likely source of this additional scatter is lack of repeatability in the daily calibrations used to establish the order locations and wavelength dispersion relations for the 2-D spectra.

\section{Discussion and Conclusions}
\label{sec:concluding}

Three years of radial-velocity monitoring the Sun as a star with the HARPS-N instrument reveal variability on timescales ranging from minutes to days to years. 

Within a single day, the radial-velocity signal is strongly correlated on timescales of minutes to hours. We find that 15-minute integrations attenuate the autocorrelation amplitude of the data by a factor 2 or so better than 5-minute sampling. The same autocorrelation analysis shows that samples taken 2 or more hours apart are effectively statistically independent. This supports the recommendations of the earlier study of the impact of stellar granulation by \cite{2011A&A...525A.140D}, whose conclusions have determined the nightly observing strategy of surveys for terrestrial-mass planets around bright stars with HARPS and HARPS-N.

Finally, the solar data reveal a need for more sophisticated data-reduction strategies. High-precision instruments such as HARPS-N require comprehensive daily calibration sequences to monitor the long-term stability of the instrument. No calibration is perfect, however, and the solar data reveal radial-velocity zero-point errors with an RMS scatter of order 0.4 m~s$^{-1}$. These arise mainly from day-to-day zero-point uncertainty in the daily wavelength calibrations carried out by the version of the pipeline with which these data were reduced. A recent major upgrade to the HARPS/HARPS-N data-reduction system has introduced a better-constrained 2-dimensional fit across all \'echelle orders into these daily calibrations for stellar data. Improvements can therefore be expected in future data releases from this project, once the 58000 existing solar spectra and future observations have been reprocessed. 

Long-term changes in the area of the CCF appear to arise from cycle-related changes in the magnetic network. There is no rotational modulation component to this signal, and the impact on the radial velocity is not yet fully clear. There appears to be no strong trend in RV over the three years of this study, but we caution that we have not yet had the opportunity to observe the full range of activity in a solar cycle.

The long-term evolution of the BIS, which measures the core-wing asymmetry of the CCF, has a baseline which mirrors the long-term evolution of the CCF area. Bipolar magnetic active regions passing across the solar disk cause perturbations lasting several days in both the radial velocity and the CCF profile parameters. The line-shape parameters of the CCF appear to respond to different components of the active regions. The FWHM shows modulation behaviour that tracks the sunspot number. The BIS shows more persistently repetitive modulation, suggesting that the line asymmetry responds primarily to disruption of the photospheric granulation pattern in the faculae, which are the longest-lived components of active regions. 

There is no single proxy indicator that can recover the RV from an isolated measurement of the line-profile shape or other activity indicators without reference to time-domain information. Given that active regions in solar-type stars evolve on timescales comparable to the rotation period, it is therefore mandatory to sample stellar radial-velocity signals often enough that active-region passages can be fully characterised in all proxy indicators. The radial velocity tracks the BIS closely, but the BIS lags the radial velocity by about 3 days. As Figs.~\ref{fig:timezoom} and \ref{fig:dcflags} show, these time lags are large enough that a simple linear decorrelation against BIS or FWHM cannot be used to correct the radial velocity for the effects of magnetic activity. A more sophisticated approach which uses time-domain information as well as line-shape information is needed. Although such a treatment is beyond the scope of this paper, we refer the reader to recent papers by \cite{2012MNRAS.419.3147A}, \cite{2015MNRAS.452.2269R} and \cite{2017arXiv171101318J}. These multivariate Gaussian-process regression approaches use temporal derivative models of RV proxy indicators in a general way that has strong potential capability to model such time lags.

\section*{Acknowledgements}

The HARPS-N project has been funded by the Prodex Program of the Swiss Space Office (SSO), the Harvard University Origins of Life Initiative (HUOLI), the Scottish Universities Physics Alliance (SUPA), the University of Geneva, the Smithsonian Astrophysical Observatory (SAO), and the Italian National Astrophysical Institute (INAF), the University of St Andrews, Queen's University Belfast,
and the University of Edinburgh.
ACC acknowledges support from the Science \&\ Technology Facilities Council (STFC) consolidated grant number ST/R000824/1 and UKSA grant ST/R003203/1, and the support of the visiting scientist program at Lowell Observatory, where part of this work was carried out. 
DP and RW acknowledge partial support by NASA award number NNX16AD42G. XD is grateful to the Branco Weiss Fellowship Society in Science for its financial support. SHS is grateful for support from NASA Heliophysics LWS grant NNX16AB79G.
CAW acknowledges support by STFC grant ST/P000312/1. The research leading to these results received funding from the
European Union Seventh Framework Programme (FP7/2007-2013) under grant agreement number 313014 (ETAEARTH). 
This work was performed in part under contract with the California Institute of Technology (Caltech)/Jet Propulsion Laboratory (JPL) funded by NASA through the Sagan Fellowship Program executed by the NASA Exoplanet Science Institute (RDH).
DWL acknowledges partial support from the Kepler mission under NASA Cooperative Agreement NNX13AB58A with the Smithsonian Astrophysical Observatory. LM acknowledges the support by INAF/Frontiera through the "Progetti Premiali" funding scheme of the Italian Ministry of Education, University, and Research. Some of this work has been carried out within the framework of the NCCR PlanetS, supported by the Swiss National Science Foundation. FP further acknowledges support by the Swiss National Science Foundation SNSF through grant number 166227.

%%%%%%%%%%%%%%%%%%%%%%%%%%%%%%%%%%%%%%%%%%%%%%%%%%



%\bibliographystyle{mnras}
%\bibliography{SILSO}



%%%%%%%%%%%%%%%%%%%%%%%%%%%%%%%%%%%%%%%%%%%%%%%%%%

%%%%%%%%%%%%%%%%% APPENDICES %%%%%%%%%%%%%%%%%%%%%

\appendix

\newpage
\section{Daily extinction and mixture model parameters}
\label{sec:daily}

\begin{table*}
\caption{Dates and daily extinction and mixture-model parameters for each day of observation. }
\label{tab:daily}
\begin{tabular}{ccccccc}
\hline\\
JD-2400000.0 & $k_{60}$ & $m_{60}(x=0)$ & $b_{\rm bg}$ & $\sigma_{\rm bg}$ & $\sigma_{\rm jit}$ & Q \\
Day & Mag/airmass & Mag & Mag & Mag & \\
\hline\\
57222.1914 & 0.1415 & -13.280 & 0.509 & 0.394 & 0.0030 & 0.931\\
57223.9179 & 0.1213 & -13.369 & 0.456 & 0.391 & 0.0085 & 0.884\\
57233.0057 & 0.0965 & -13.323 & 0.734 & 0.386 & 0.0035 & 0.976\\
57234.0432 & 0.0891 & -13.309 & 0.462 & 0.418 & 0.0042 & 0.991\\
57235.0145 & 0.0995 & -13.326 & 0.465 & 0.425 & 0.0044 & 0.993\\
57235.9870 & 0.1074 & -13.335 & 0.471 & 0.391 & 0.0059 & 0.991\\
57237.0092 & 0.1089 & -13.346 & 0.489 & 0.422 & 0.0058 & 0.993\\
57237.9919 & 0.0902 & -13.262 & 0.506 & 0.414 & 0.0103 & 0.991\\
57239.0619 & 0.1351 & -13.339 & 0.479 & 0.383 & 0.0016 & 0.979\\
57239.9333 & 0.1090 & -13.321 & 0.623 & 0.354 & 0.0065 & 0.965\\
57242.0245 & 0.2982 & -13.463 & 0.432 & 0.417 & 0.0505 & 0.982\\
57242.8892 & 0.3109 & -13.355 & 0.335 & 0.564 & 0.0561 & 0.330\\
57244.9912 & 0.2314 & -13.353 & 0.181 & 0.290 & 0.0058 & 0.670\\
57252.0397 & 0.0963 & -13.320 & 0.634 & 0.534 & 0.0079 & 0.976\\
57252.9576 & 0.0841 & -13.297 & 0.437 & 0.547 & 0.0079 & 0.742\\
57253.9311 & 0.0829 & -13.297 & 0.882 & 0.305 & 0.0071 & 0.944\\
.&.&.&.&.&.&.\\
.&.&.&.&.&.&.\\
.&.&.&.&.&.&.\\
\hline\\
\end{tabular}
\end{table*}

Table~\ref{tab:daily} lists Julian dates and daily extinction and mixture-model parameters for each day of observation. The Julian date is the median of the observation dates on the day to which each record applies. The values of the extinction coefficient $k_{60}$, the magnitude zero point $m_{60}(x=0)$ above the atmosphere, the background population mean magnitude offset $b_{\rm bg}$ and standard deviation $\sigma_{\rm bg}$, the transparency fluctuation amplitude $\sigma_{\rm jit}$ and the prior mixture fraction $Q$ are the medians of the final MCMC samples computed for each day. Only the first sixteen days of the campaign are listed here. The full table is available online-only, in machine-readable form as a comma-separated variable (CSV) file.

\newpage
\section{Radial velocities and CCF profile parameters}
\label{sec:alldata}

\begin{table*}
\caption{Radial velocities and CCF profile parameters (provided online-only). }
\label{tab:alldata}
{\tiny
\setlength\tabcolsep{2pt} % default value: 6pt
\begin{tabular}{cccccccccccccccccc}
\hline\\
BJD&JD&RV$_{\rm bary}$&RV$_{\rm hel}$&RV$_{\rm final}$&RV$_{\rm err}$&FWHM$_{\rm obs}$&FWHM$_{\rm sid}$&Contrast&BIS$_{\rm obs}$&H&Airmass&$m_{60}$&$\sigma(m_{60})$&Dec&Diam&PA&Qualflag\\
\hline\\
57222.1780 & 57222.1838 & 0.1122 & 0.1040 & 0.1043 & 0.0004 & 7.0064 & 7.0485 & 45.990 & -0.0186 & 0.820 & 1.366 & -13.0887 & 0.0026 & 0.36708 & 0.009155 & 0.0886 & 0.9992\\
57222.1817 & 57222.1876 & 0.1123 & 0.1041 & 0.1044 & 0.0004 & 7.0081 & 7.0502 & 45.987 & -0.0193 & 0.843 & 1.393 & -13.0840 & 0.0026 & 0.36706 & 0.009155 & 0.0886 & 0.9993\\
57222.1855 & 57222.1913 & 0.1121 & 0.1040 & 0.1043 & 0.0004 & 7.0066 & 7.0487 & 45.992 & -0.0184 & 0.867 & 1.422 & -13.0750 & 0.0026 & 0.36705 & 0.009155 & 0.0886 & 0.9990\\
57222.1892 & 57222.1951 & 0.1115 & 0.1034 & 0.1037 & 0.0004 & 7.0066 & 7.0488 & 45.999 & -0.0174 & 0.890 & 1.453 & -13.0713 & 0.0026 & 0.36704 & 0.009155 & 0.0887 & 0.9991\\
57222.1930 & 57222.1989 & 0.1119 & 0.1038 & 0.1041 & 0.0004 & 7.0083 & 7.0504 & 45.989 & -0.0194 & 0.914 & 1.485 & -13.0745 & 0.0026 & 0.36703 & 0.009155 & 0.0887 & 0.9985\\
.&.&.&.&.&.&.&.&.&.&.&.&.&.&.&.&.&.\\
.&.&.&.&.&.&.&.&.&.&.&.&.&.&.&.&.&.\\
.&.&.&.&.&.&.&.&.&.&.&.&.&.&.&.&.&.\\
\hline\\
\end{tabular}
}
\end{table*}

Table~\ref{tab:alldata} provides the full set of radial velocities and CCF profile parameters. It is provided online-only in machine-readable form as a CSV file. The columns are:\\
(1) Barycentric JD; \\
(2) JD; \\
(3) Barycentric RV in km~s$^{-1}$;\\
(4) Heliocentric RV in km~s$^{-1}$;\\
(5) Heliocentric RV in km~s$^{-1}$, corrected for differential extinction;\\
(6) RV error in km~s$^{-1}$;\\
(7) Observed CCF FWHM in km~s$^{-1}$;\\
(8) Sidereal CCF FWHM  in km~s$^{-1}$, corrected for Earth orbital motion and solar obliquity; \\
(9) CCF contrast (percent);\\
(10) Observed CCF BIS in km~s$^{-1}$;\\
(11) Apparent hour angle in radians;\\
(12) Apparent Airmass;\\
(13) Pseudo-magnitude derived from SNR in order 60;\\
(14) Magnitude error derived from SNR in order 60;\\
(15) Solar declination $\delta$ (radian);\\
(16) Solar angular diameter (radian);\\
(17) Position angle of north solar rotation pole (radian);\\
(18) Mixture-model foreground membership probability $p(q_i=0)$.

%%%%%%%%%%%%%%%%%%%%%%%%%%%%%%%%%%%%%%%%%%%%%%%%%%

% Don't change these lines
\bsp	% typesetting comment
\label{lastpage}
\end{document}